**Requirements for beneficial electrochemical restructuring: A model study on a cobalt oxide in selected electrolytes**

*Javier Villalobos[1], Diego González-Flores[2,3,*], Roberto Urcuyo[2,3], Mavis L. Montero[2], Götz Schuck[4], Paul Beyer[5], Marcel Risch[1,*]*

1. J. Villalobos, Dr. M. Risch
Nachwuchsgruppe Gestaltung des Sauerstoffentwicklungsmechanismus, Helmholtz-Zentrum Berlin für Materialien und Energie GmbH, Hahn-Meitner Platz 1, 14109, Berlin, Germany
E-mail: marcel.risch@helmholtz-berlin.de

2. Dr. D. Gonzalez-Flores, Dr. R. Urcuyo, Dr. M. L. Montero
Centro de Investigación en Ciencia e Ingeniería de Materiales (CICIMA), San José 11501 2060, Costa Rica
Escuela de Química, Universidad de Costa Rica, San José 11501 2060, Costa Rica
E-mail: diegoandres.gonzalez@ucr.ac.cr

3. Dr. D. Gonzalez-Flores, Dr. R. Urcuyo
Centro de Electroquímica y Energía Química (CELEQ), San José 11501 2060, Costa Rica

4. Dr. G. Schuck
Abteilung Struktur und Dynamik von Energiematerialien, Helmholtz-Zentrum Berlin für Materialien und Energie GmbH, Berlin 14109, Germany

5. Dr. P. Beyer
Fachbereich Physik, Freie Universität Berlin, Arnimallee 14, 14195 Berlin, Germany

The requirements for beneficial materials restructuring into a higher performance OER electrocatalyst are still a largely open question. Here we use Erythrite ($Co_3(AsO_4)_2{\cdot}8H_2O$) as a Co-based OER electrocatalyst to evaluate its catalytic properties during in-situ restructuring into an amorphous Co-based catalyst in four different electrolytes at pH 7. Using diffraction, microscopy and spectroscopy, we observed a strong effect in the restructuring kinetics depending of the anions in the electrolyte. Only carbonate electrolyte could activate the catalyst electrode, which we relate to its slow restructuring kinetics. While its turnover frequency (TOF) reduced from 2.84 $O_2$ $Co^{-1}$ $s^{-1}$ to a constant value of 0.10 $O_2$ $Co^{-1}$ $s^{-1}$ after ~ 300 cycles, the number of redox active sites continuously increased, which explained the current increase of around 100%. The final activated material owns an adequate local order, a high Co oxidation state and a high number of redox-active Co ions, which we identify as the trinity for enhancing the OER activity. Thus, this work provides new insights into for the rational design of high-performance OER catalysts by electrochemical restructuring.

## 1. Introduction

The widespread transition use of renewable energy requires efficient energy storage solutions due to the fluctuating energy production from renewable sources such as sunlight or wind. A promising solution is the storage of chemical energy by water splitting into hydrogen and oxygen.[1,2] Complex kinetics makes the oxygen evolution reaction (OER) one of the most considerable challenges for implementing water splitting as it requires high catalytic efficiency and stability under operating conditions.[3] For the use of carbon-based fuels as energy storage, near-neutral pH operating conditions are desirable to couple the anodic OER to the cathodic $CO_2$ reduction reaction since $CO_2$-enrichment of the electrolyte leads to pH values close to 7.[4,5] Amorphous transition-metal oxides have shown outstanding catalytic properties at neutral pH.[6–8] Thanks to their unique atomic arrangement, these materials own structural flexibility and distinctive coordinated metal center.[9–11] Electrochemically, amorphous oxides can be obtained by electrodeposition[12–14] or electrochemical restructuring[9,15–19] of crystalline materials. The latter approach has improved catalytic activity compared to their crystalline variant in some cases,[9,10,12,20] while this was not the case for other combinations of pristine materials and electrolytes.[21,22] Wang et al.[23] identified an optimal Co redox level for the activation of $LiCoO_{2-x}Cl_x$ by beneficial electrochemical restructuring in 1 M KOH. The requirements for beneficial electrochemical restructuring, particularly in neutral electrolytes, are still largely an open question.[18,24]

Among a wide range of amorphous transition-metal oxides, Co-containing oxides own particular properties, such as self-healing,[25,26] semiconductivity,[27–29] and the possibility to host higher oxidation states, such as Co(IV),[8,30] which has been proposed as part of the active site.[31] Co-based oxides present overpotentials of 400-490 mV at 10 mA $cm^{-2}$, or 260 mV at 1 mA $cm^{-2}$ and Tafel slopes of 60-80 mV $dec^{-1}$, in near-neutral pH.[31–34] The local structure of the Co oxide differs depending on the cations and anions present in the electrodeposition electrolyte; for instance, the local order of the final Co oxide decreases as a function of the electrolyte as it follows $CaCl_2$>KCl>LiOAc>KOAc>KPi.[8] The local order parameter can be used to estimate the size of the Co oxide fragments (clusters) using X-ray absorption spectroscopy,[35] the authors suggested an optimal local order parameter, where the catalytic current was highest in their study. Additionally, Kwon et al.[36] reported the use of different anions into the electrodeposition electrolyte of Co oxide, resulting in larger interlayer space and cluster size as follows: borate (pH 9.2) > methyl phosphate (pH 8.0) > phosphate (pH 7.0). Thus, crystalline Co oxides are attractive starting materials for further enhancement by electrochemical restructuring into more active amorphous oxides.

In this study, we synthesized Erythrite ($Co_3(AsO_4)_2 \cdot 8H_2O$) as a Co-based catalyst model to evaluate its catalytic properties during its electrochemical restructuring into partially amorphous Co oxide in four different electrolytes at pH 7. The three electrolytes with clearly different restructuring behavior were chosen to track the conversion during cyclic voltammetry by diffraction, microscopy, and spectroscopy. We observed that the kinetics of the electrochemical restructuring strongly depend on the electrolyte anion. Only carbonate electrolyte can activate catalytic current over cycling due to slow restructuring kinetics. The final activated material shows an adequate cluster size, a high Co oxidation state, and high Co redox activity, which are essential features enhancing the catalytic activity. Hence, this work provides new insights into the requirements of beneficial electrochemical restructuring of Co-based materials.

## 2. Results and discussion

### 2.1. Electrochemical restructuring of Ery in selected electrolytes

Erythrite (Ery; $Co_3(AsO_4)_2 \cdot 8H_2O$) was chosen as a model system to study the effect of electrochemical restructuring on catalytic activity in selected neutral electrolytes with different anions. Pristine Ery is a crystalline material, part of a group of isostructural minerals called vivianites.[37] We synthesized it by a thermal process at 65 °C, which assures crystal growth. Physical and chemical characterization was carried out to confirm the formation of the Ery phase (**Figure S1**). In the chemical structure of Ery,[38,39] arsenate anions bind two positions of Co atoms: Co(1) on the hexacoordinated position and Co(2) bound via di-μ-oxo(arsenate)-bridge (**Figure 1**; various structures in **Figure S2**). Some of us have previously used Ery[40] and other isostructural materials[12] as catalysts for OER. We have observed their trend to lose long-range order (commonly called amorphization) after some catalytic cycles under OER conditions, i.e., electrochemical restructuring. This transformation promoted layered cobalt oxide formation, which has significantly different interatomic distances (**Figure 1**). For instance, Co(2) – Co(2) distance changed from 3.1 Å in Ery to about 2.8 Å after restructuring **(Figure 1**).

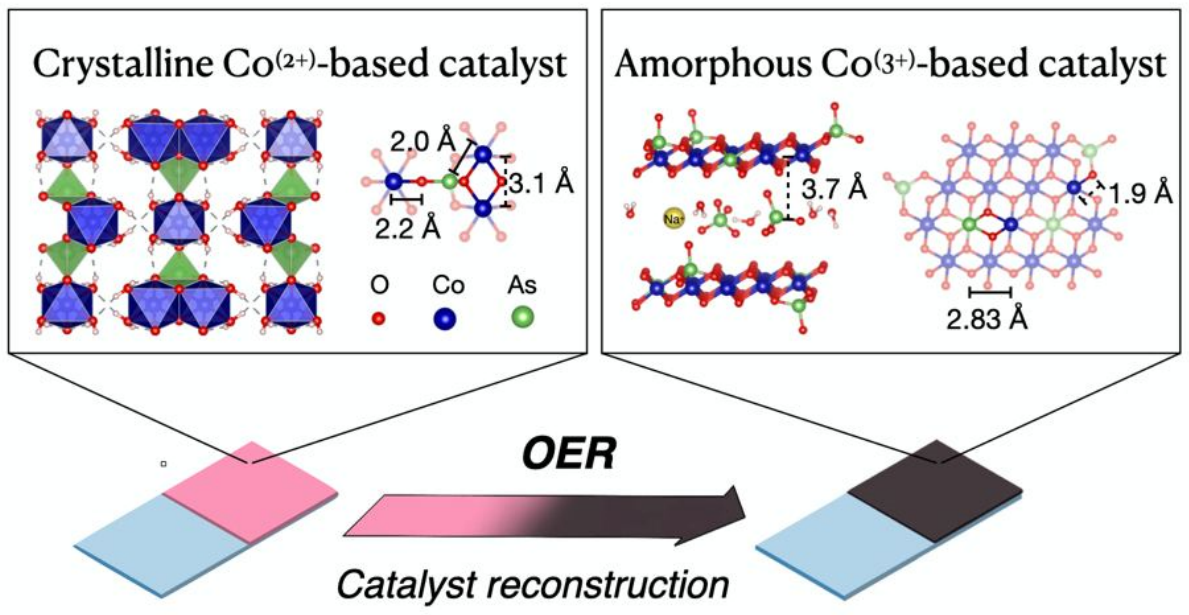

**Figure 1.** Diagram of the electrochemical restructuring of Ery. The crystalline Ery ($Co_3(AsO_4)_2 \cdot 8H_2O$) layers are joined by hydrogen bonds. Arsenate binding in Ery, Co – As, and Co – Co distances are shown. The amorphous layered Co oxide structure, distances of different binding modes from arsenate anions into the layer and at the border are shown. Ions and water molecules are accumulated in the interlayer space, a typical layer distance is displayed.

Ery deposited on FTO glass showed clearly different trends during cyclic voltammetry (CV) in four different electrolytes at pH 7 (electrochemical protocol, **Table S1**). The used anions were 0.1 M of borate, phosphate, carbonate, and arsenate, resulting in the restructured catalysts Ery-BO3, Ery-PO4, Ery-CO3, and Ery-AsO4, respectively (**Figure 2** and **Figure S3**). A high number of cycles (here 800 cycles), a high upper potential limit (2.1 V vs. RHE), and a high sweep speed (100 $mVs^{-1}$) were used as typical for stability, restructuring, or activation studies.[14,41–44] The maximum current, $i_{max}$ (at 2.1 V vs. RHE), was comparable among all different electrolytes during the first cycles, i.e., it is mainly that of the as-synthesized Ery. Yet, $i_{max}$ continuously increased in carbonate electrolyte, whereas, in phosphate and borate electrolyte, it rises during the first 200-300 cycles, after which it started decreasing. No significant $i_{max}$ changes were observed in arsenate electrolyte, indicating an equilibrium with arsenate in Ery and the electrolyte that prevents significant catalytic changes. All trends were reproducible in three trials (**Figure S4**).

In addition to catalytic changes, increased broad redox peaks were observed with cycling, which has previously been associated with the number of redox-active metal ions[8,13] and in crystalline materials with the electrochemical restructuring to an amorphous material.[9,12,40] The redox-active metal ions can be quantified by integrating the reductive currents in the CV resulting in an electro-redox charge (ERC). In Co-based catalysts, the ERC correlates with the redox activity of Co during the catalytic reaction, as detailed elsewhere.[8,12,13] ERC trends over cycling were estimated in different electrolytes (**Figure S5**). Ery-BO3 reached a steady-state at cycle 300 (ERC = 11.4 mC), whereas Ery-PO4 and Ery-CO3 increased constantly over 800 cycles and did not reach a steady-state; the final ERC values are 15.5 mC and 16.0 mC, respectively. Similarly to the activity trends, the changes for Ery-AsO4 were small (ERC=3.3 mC, after 800 cycles).

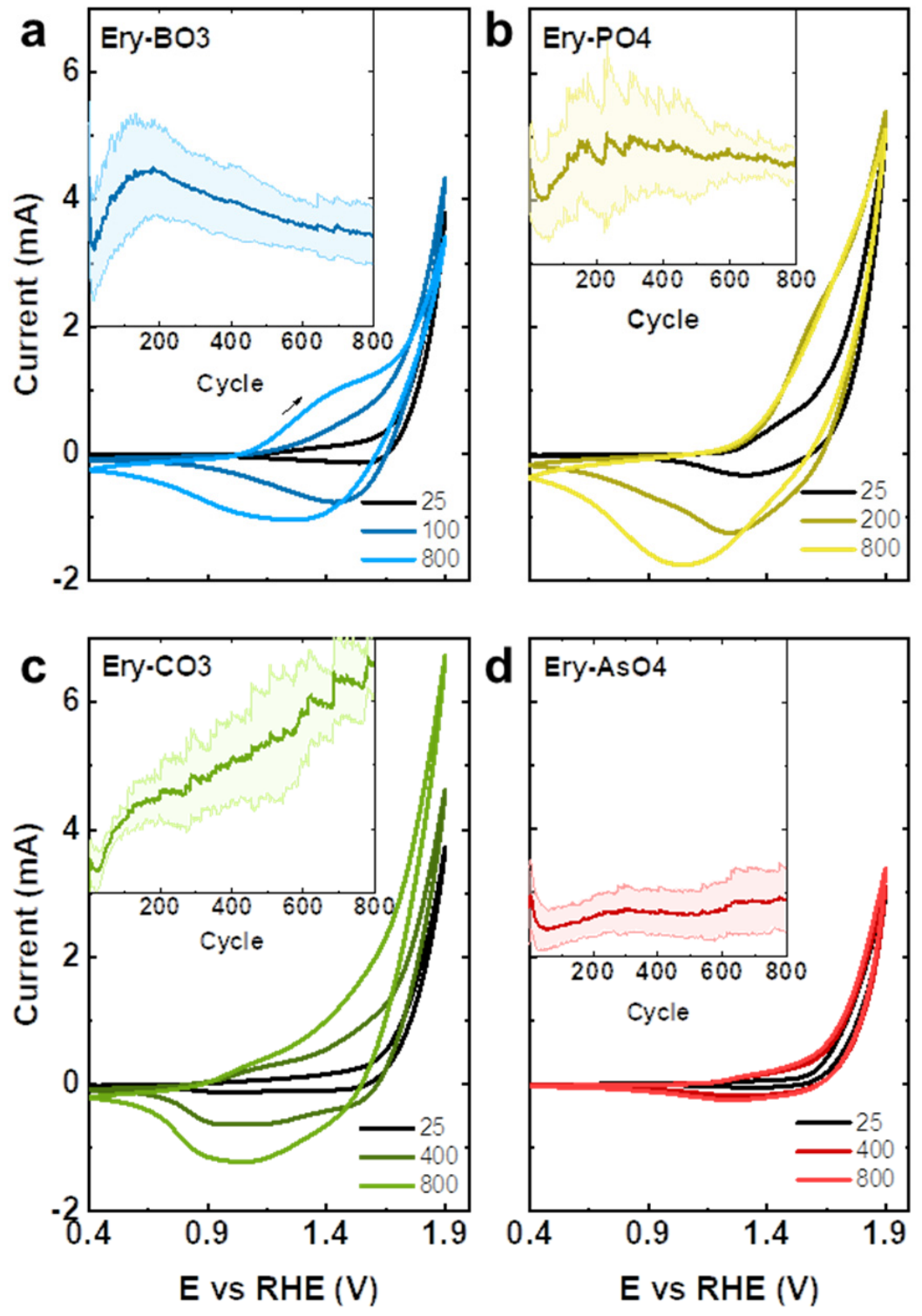

**Figure 2.** Series of CV performed on an Erythrite-deposited FTO glass in different electrolytes at pH 7 and a concentration of 0.1 M: a. borate, b. phosphate, c. carbonate, and d. arsenate. 800 cycles were performed with a sweep speed of 100 mV $s^{-1}$ and 85 % iR compensation. Potential is represented as E vs. RHE. The arrows show the direction of the CV. The insets show the $i_{max}$ trends (at E=2.1 V vs. RHE) as a function of cycling. The light-colored area represents the standard deviation of three different samples.

The potential of the maximum of the reductive (i.e., cathodic) peak significantly differed depending on the electrolyte. The initial peak potentials were around 1.45 V vs. RHE in the four electrolytes, i.e., that of pristine Ery. They changed to 1.28 V vs. RHE in Ery-PO4, 1.57 V vs. RHE in Ery-BO3, 1.24 V vs. RHE in Ery-CO3, and 1.47 V vs. RHE in Ery-AsO4. Risch et al.[31] reported similar redox peak positions for CoCat (abbreviation for Co-catalyst, electrodeposited in phosphate buffer, KPi, pH 7). They assigned the midpoint potential between the anodic and cathodic redox peaks at 1.42 V vs. RHE to the $Co^{2+}/Co^{3+}$ transition and at 1.63 V vs. RHE to the $Co^{3+}/Co^{4+}$ transition. The latter is not clearly resolved in our CV series (**Figure 2**). Similar peak positions were reported by Villalobos et al.[40], who also showed that the use of arsenate electrolyte for the electrodeposition of CoCat (instead of KPi) resulted in a shift of 0.1 V in the peak position of the $Co^{2+}/Co^{3+}$ transition to lower potential, whereas the $Co^{3+}/Co^{4+}$ remained unaffected. Therefore, the position of the waves in the CV depends on the electrolyte used for restructuring, which means that the energy level of the Co redox can be tuned simply by selection of the electrolyte anion similarly to the approach of Wang et al.[23] who tune the cation of the catalyst material.

Three indicators were used to monitor the electrochemical restructuring during cycling in the three electrolytes with the most pronounced change, namely borate, phosphate and carbonate. The indicators were: (I) estimation of the number of redox-active metal ions by the quantification of ERC, (II) loss of the Ery phase from the initial material by XRD using the normalized (020) reflection of Ery (**Figure S6**), and (III) anion exchange tracked as loss of arsenate from Ery ($Co_3(AsO_4)_2{\cdot}8H_2O$) by EDX **(Figure S7**). The loss of arsenate anions was also tracked by the signal of the As-O bond[45] at 780 $cm^{-1}$ in FTIR (**Figure S8**), corroborating the EDX results. These tracking experiments were based on the methods and properties most commonly used to understand the electrochemical restructuring process.[10,12,15,40,46] How the changes of metal redox, loss of crystallinity, and anionic exchange relate to each other has been an important open question for understanding the mechanism of electrochemical restructuring. In this regard, these three indicators were compared as a function of cycling in borate, carbonate, and phosphate electrolytes to identify trends and correlations (**Figure 3**). They were complemented by $i_{max}$ as an indicator of activity.

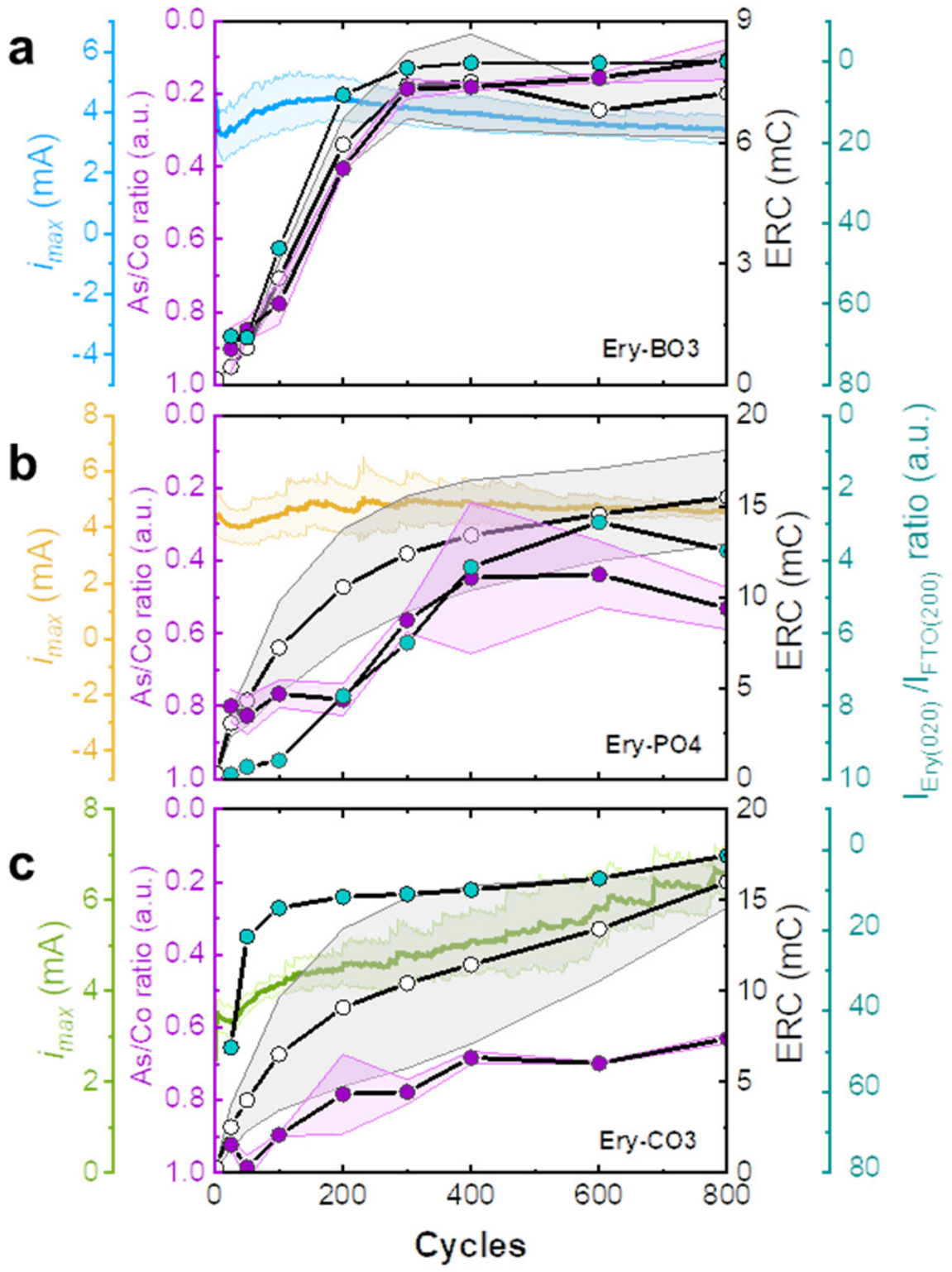

**Figure 3.** As/Co ratio, $I_{Ery(020)}/I_{FTO(200)}$, and ERC as a function of cycling for Ery in: **a.** borate electrolyte, **b.** phosphate electrolyte and **c.** Carbonate electrolyte. The light-colored area represents the standard deviation of three different measurements.

In Ery-BO3 (**Figure 3a**), a plateau was reached by cycle 300 in all indicators, where the material has lost most of Ery's arsenate anions (Co/As ≈ 80 %), the ERC reached a value of only 7 mC, and it did not show extended long-range order anymore, i.e., the (020) reflection is < 1% of its initial value. Thus, 300 cycles in borate electrolyte were sufficient to conclude the electrochemical restructuring. SEM images taken at selected cycles showed a constant homogeneous coverage of the substrate for all cycles (**Figure S9**). However, the initial needle-like morphology melted with cycling, which was most pronounced after the $300^{th}$ cycle (**Figure S9e**). Interestingly, the activity measured by $i_{max}$ reached its maximum at cycle 180 before the restructuring was completed and then started to decrease, indicating the additional relevance of further materials changes, such as the morphology, for catalytic activity. In summary, all restructuring indicators were correlated for Ery-BO3.

In Ery-PO4 (**Figure 3b**), the ERC trend differed from the other two indicators of the As/Co ratio and the loss of the Ery(020) reflection. These two indicators shared the same trend and mainly increase during the first 400 cycles. (Note that their axes are plotted with the lowest value on top to better compare them to the ERC). Less than 50 % of the arsenate was lost in Ery-PO4, and the material still showed sizable long-range order after 800 cycles, meaning that the processes follow different kinetics in phosphate electrolytes as compared to borate electrolytes. Furthermore, the increase in redox activity (i.e., ERC) does not solely depend on restructuring the material. The ERC increased continuously over 800 cycles and reached a final value of 15.5 mC; an ERC increase has commonly been related to the generation of more active sites.[8,12,13] SEM images showed full coverage again at all cycles and a needle-like morphology but have no clear trend with cycling (**Figure S10**). The activity indicator, $i_{max}$, did not change remarkably over cycling, yet a broad maximum of the average $i_{max}$ was found around cycle 300, again before the restructuring was completed. In summary, for Ery-PO4, the loss of the Ery crystalline structure correlates with the As loss, whereas the ERC differs, rising continuously.

In Ery-CO3, the ERC continuously increased continuously, whereas the As/Co ratio and the Ery crystallinity mainly decreased during the first 200 cycles and then remained nearly constant. On this plateau, less than 50 % of the arsenate was lost. After 200 cycles, the Ery crystallinity practically vanished ($I_{Ery(020)}/I_{FTO(200)} \approx 10$) whereas no crystallinity could be detected after 800 cycles ($I_{Ery(020)}/I_{FTO(200)} < 1$, **Figure S11**). In contrast, the ERC increased continuously over cycling and reached a value of 16 mC (the highest among the three electrolytes). Therefore, the As/Co ratio decreases and crystallinity loss followed different kinetics in Ery-CO3, besides the different kinetics of ERC, which was already observed in Ery-PO4. Interestingly, the activity indicator $i_{max}$ decreased during the first 25 cycles, yet it increased continuously from cycle 25 up to cycle 800, indicating an evident activation of around 100 % in the current of Ery-CO3. In summary, none of the restructuring indicators correlated for Ery-CO3, yet the electrochemical current increased monotonously.

Taken together, the data in **Figure 3** showed that the electrolyte anions affected the restructuring kinetics of Ery (illustrated as reaction order fitting in **Figure S12** and **Table S2**) and clearly influenced how the different dynamic processes, e.g., loss of Ery crystallinity, anionic exchange, and increase of redox-active Co (i.e., ERC), correlate with each other. Previous reports have shown the relation between the electrolyte pH[47] or the cation stoichiometry[23] with the efficiency of restructuring. Here, the lack of correlation between the ERC and the other restructuring indicators for some electrolytes suggests a further change in the resulting local and electronic structures as additional factors.

We used XAS to understand the changes in local structure during cycling in different electrolytes as it does not require crystallinity and has high chemical sensitivity. The Co-K edge was used to analyze the Co oxidation state and local structure after a selected number of cycles. The nominal Co oxidation state was estimated by calibration with three Co references in different oxidation states (**Figure S13** and **Table S3**). The XANES spectrum of pristine Ery had a Co oxidation state of 2+ (**Figure 4a**). For Ery-BO3, the catalyst material showed a shift in the X-ray edge position to higher energy, indicating oxidation to 2.3+ at 100 cycles and 2.8+ at 800 cycles. Similar continuous oxidation was observed for Ery-CO3 from 2+ to 2.4+ at 100 cycles and 2.8+ at 800 cycles. The Co atoms of Ery-PO4 oxidized to 2.4+ at cycle 100, which remained unchanged until cycle 800. A double maximum (white line) in the XANES spectra suggested a combination of two different phases, Ery and another Co oxide,[40] to be identified below. It was observed in all spectra for Ery-PO4. In Ery-BO3, the double white line was observed only in cycle $100^{th}$ but not in cycle $800^{th}$, indicating a single-phase material in agreement with the vanished Ery (020) reflection in XRD. In Ery-CO3, the double white line was observed only after 800 cycles since at cycle 100, the other Co oxide contributed negligibly.

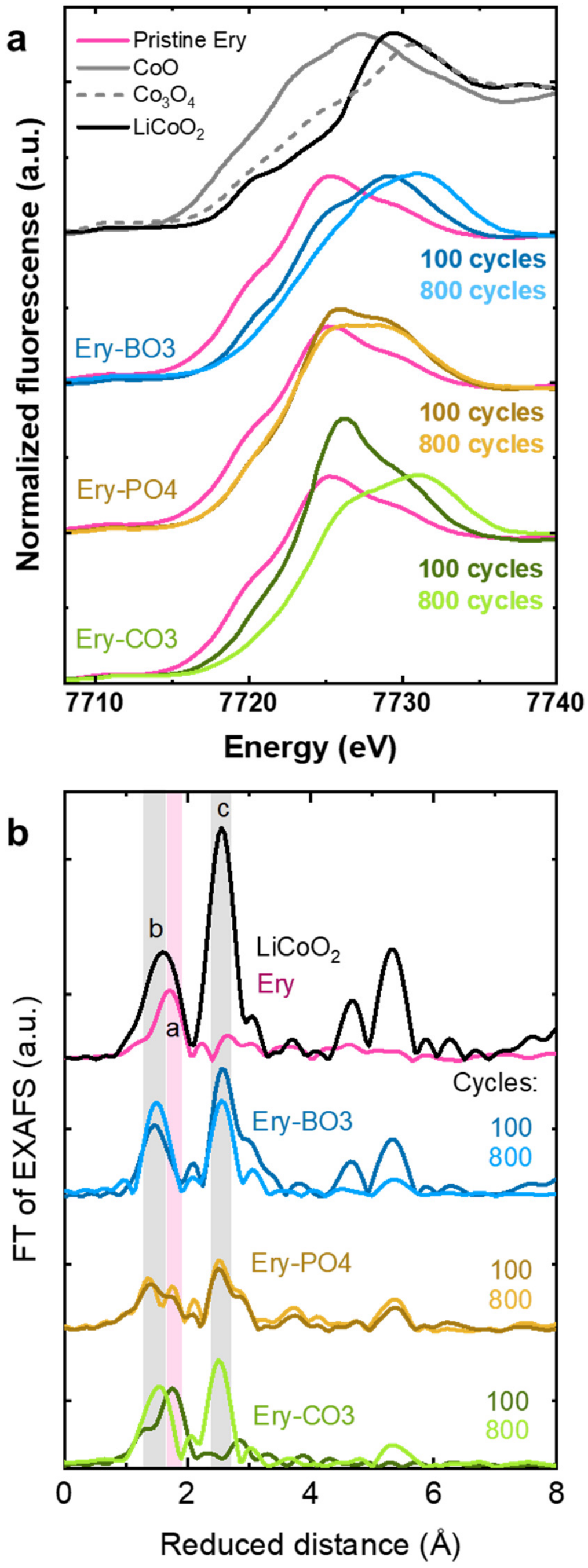

**Figure 4.** a. XANES spectra of Co-K edge collected on Ery after selected cycles in **a.** pH 7 0.1 M borate electrolyte and **b.** pH 7 0.1 M phosphate electrolyte. The Co-K edge spectra of $Co^{(2+)}O$, $Co_3^{(2.7+)}O_4$, and $LiCo^{(3+)}O_2$ were added as references, b. Fourier-transform EXAFS spectra of Co-K edge collected on Ery after selected cycles in: pH 7 0.1 M borate electrolyte (blue lines), pH 7 0.1 M phosphate electrolyte (brown-yellow lines) and pH 0.1 M carbonate electrolyte (green lines). The light-colored gray areas highlight the interatomic distances related to Co oxide (peak b and c), whereas the light-colored pink areas show the Ery-typical distance of 2.1 Å (Co-O, peak a). The reduced distance is about 0.3 Å shorter than the precise distance obtained by EXAFS simulations (**Figure S14-S16, Table S4**).

The Co-K edge of the Fourier transform (FT) of the EXAFS was used to study the local structure of the electrochemically restructured Ery after a selected number of cycles in borate, electrolyte, and carbonate (**Figure 4b**). The FT of pristine Ery has a prominent peak at 2.1 Å (Co-O) and a weaker one at 3.1 Å (Co-Co).[40]

The spectrum of Ery-PO4 showed EXAFS peaks similar to pristine Ery after 100 and 800 cycles having the above Co-O and Co-Co distances typical of Ery.[40] Additionally, distances at 1.9 Å (Co-O) and 2.85 Å (Co-Co) were visible; these are distinct signals of layered Co oxides such as $LiCoO_2$ that is also shown in **Figure 4b** for reference. The lack of change in the spectra was expected based on the XANES analysis above. The presence of Ery and a layered Co oxide corroborates our interpretation of the double white line.

The spectrum of Ery-BO3 after 100 cycles also showed the peaks of Ery and the layered Co oxide. After cycle 800, the spectrum only showed the prominent peaks of a Co oxide, such as 1.9 Å (Co-O) and 2.8 Å (Co-Co), indicating complete electrochemical restructuring. Interestingly, the distance at 3.7 Å (Co-O, **Figure 1**) significantly decreases its amplitude. This signal has been previously associated with the distance between Co atoms into the oxide layer and oxygen atoms, either from water molecules or electrolyte ions, in the interlayer.[8,48,49] A reduction of this signal amplitude may reflect a decrease in the long-range order of the layered Co oxide.

In Ery-CO3, the complete electrochemical restructuring from Ery into the Co oxide was apparent. After the $100^{th}$ cycle, the interatomic distance at 2.1 Å (Co-O) evidenced the presence of Ery; no Co oxides distances were observed. Nonetheless, after cycle 800, the spectrum showed the interatomic distances at 1.9 Å (Co-O) and 2.85 Å (Co-Co) of the Co oxide, and signals of Ery were not clearly observed.

Since the layered Co oxide formed to various degrees in all electrolytes, we will take a closer look at its properties to rationalize the differences in activity among the restructured Ery catalysts. Amorphous Co oxides are organized in layers of several hexa-oxo-coordinated Co atoms; the number of Co atoms determines the extent of the layer fragment or cluster.[36,49,50] The cluster size of amorphous Co oxide has been previously used to understand activity trends.[8,35,48] This cluster size of the Co oxide is effectively monitored by the ratio of the Co-Co peak height to the Co-O peak height (cobalt peak ratio, CPR) in the FT of EXAFS,[35] where a large ratio indicates a large cluster or ultimately a crystalline solid such as $LiCoO_2$ in **Figure 4b**.

We plotted the activity indicator, $i_{max}$, as a function of the CPR (**Figure 5a**) to further investigate the nature of the layered Co oxide formed by restructuring Ery. It must be noted that the CPR depends on the parameters used to perform the FT, which were identical for our analysis. For each individual dataset, $i_{max}$ increased with the CPR. This explained the activity decrease with cycling for Ery-PO4 and Ery-BO3 where CPR and activity were maximal after 400 and 100 cycles. However, the exact value of the CPR does not result in the same activity for electrochemical restructuring in different electrolytes, which suggested that another parameter also strongly affected activity.

The role of a high Co oxidation state (3+ and even, 4+) has been widely reported as essential for oxygen evolution in neutral electrolytes.[13,31,51–55] Thus, we plotted the restructured Ery catalysts as function of the average Co oxidation state (**Figure 5b**) at fixed CPR of 1.3. The plot was further enriched with the ERC at each point to discuss the number of potentially active sites provided by the ERC as well as the efficiency of a single active site, likely given by its Co oxidation state.

In **Figure 5b**, after 100 cycles in phosphate, the Co oxidation of Ery-PO4 is also 2.4+, yet the ERC is only 7.2 mC, resulting in a lower catalytic current (4.4 mA) due to comparably fewer (redox) active sites as compared to Ery-PO4 after 800 cycles. After 25 cycles in borate, the ERC (1.3 mC) and Co oxidation state (2.3+) of Ery-BO3 are the lowest among the samples, associated with the lowest current (3.5 mA). In particular the comparison of Ery-PO4 at different cycles and fixed CPR clearly highlighted the importance of the number of redox active sites for high current.

For Ery-CO3, Co atoms have the highest oxidation state (2.8+), the highest ERC (16.0 mC), which results in the highest current (6.6 mA). After 800 cycles on Ery-PO4, the ERC is almost as high as that of Ery-CO3 (15.5 mC), yet the oxidation state is only 2.4+, resulting in a significantly lower catalytic current (4.6 mA). Even though the redox activity is very similar and the Co oxide clusters have an appropriate size (CPR of 1.3), a high Co oxidation state is required additionally to increase the catalytic current.[53,54,56] Therefore, we conclude that the more oxidized active sites in Ery-CO3 are more efficient as compared to Ery-PO4.

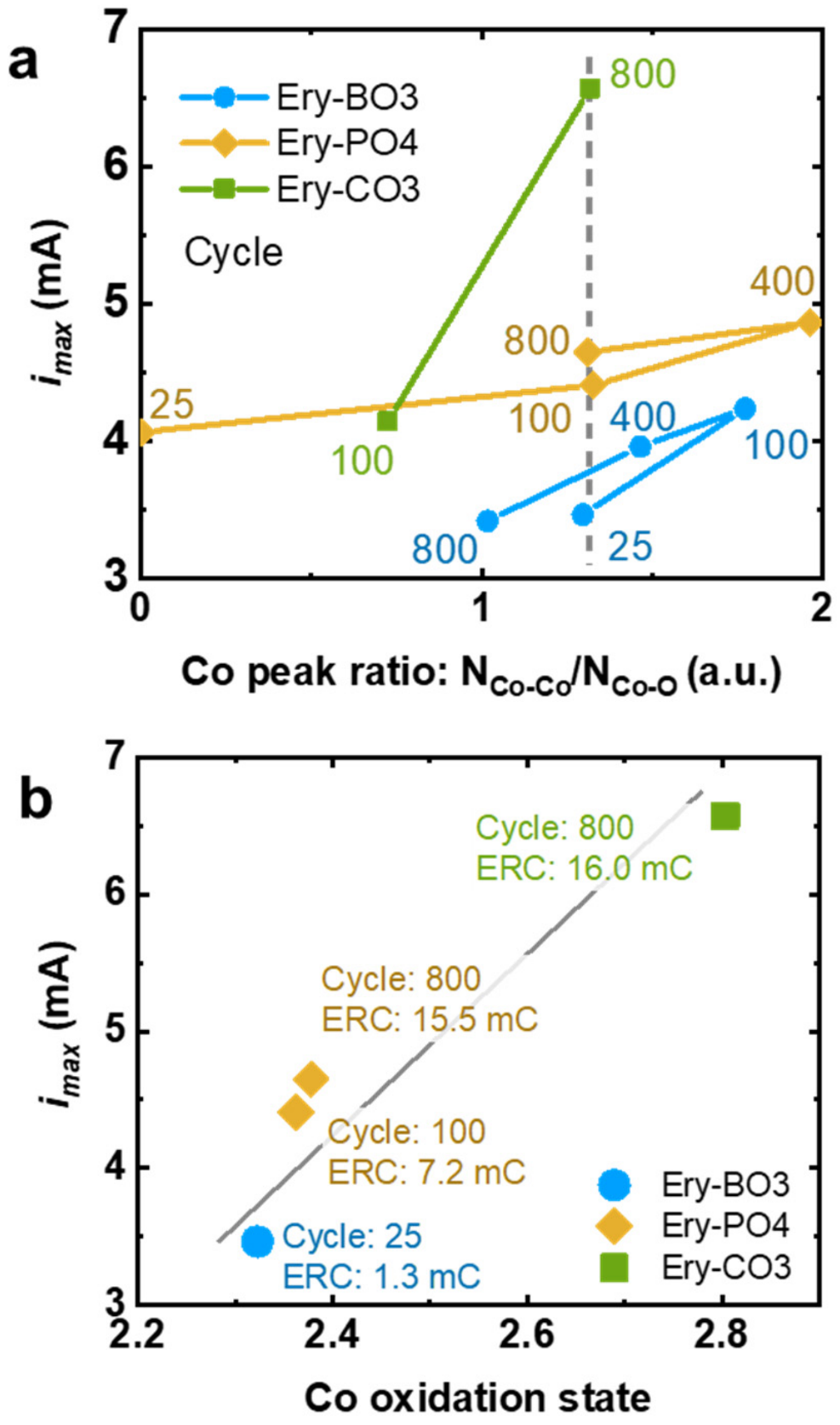

**Figure 5. a.** The activity indicator, $i_{max}$, as a function of the FT of EXAFS height peak ratio of Ery in borate, phosphate, and carbonate after a selected number of cycles (shown in graph). The solid lines bind the dots in chronological order (based on cycling), **b.** $i_{max}$ (at 2.1 V vs. RHE) as a function of the average Co oxidation state of Ery in borate (25 cycles), phosphate (100 and 800 cycles), and carbonate (800 cycles).

The efficiency of the active sites was quantified by the normalization of $i_{max}$ by ERC, which corresponds to four times (due to the four electrons transferred in the proton-coupled electron reaction) the turnover frequency of $O_2$ per Co active site (TOF ).[8,12,13] Note that ERC was used in the calculation (full detail in **Table S5**), which is an estimation of the number Co atoms with oxidation states changes; therefore it also counts sub-surface Co atoms that were oxidized and not involved in the catalytic reaction. Thus, the TOF is underestimated. The initial TOF of Ery was clearly higher than that of the restructured catalysts, as discussed in detail for a related structure,[12] suggesting that crystallinity benefits the efficiency of an active site. For Ery-PO4, the TOF decreased continuously, whereas it reached a steady value after 200-300 cycles for Ery-BO3 and Ery-CO3. The TOF of Ery-CO3 after 800 cycles ($1.3x10^{-2}$ $O_2$ $Co^{-1}$ $s^{-1}$ at 0.3 V overpotential; **Table S6**; **Figure S17**) was slightly higher than that of other Co oxides at the same overpotential, even those measured in alkaline media, where the activity is usually higher.[57,58] Moreover, its TOF was in the range of amorphous electrodeposited Co oxide,[59] supporting the that a similar surface is produced by restructuring Ery and electrodepositing Co oxide. Tafel slopes of Ery-CO3 were evaluated before and after activation in carbonate electrolyte (**Figure S18**) to examine if the current changes were due to different mechanism paths.[60] Yet, the Tafel slopes remain unaffected after activation, suggesting an unchanged mechanism. However, the Tafel slope values of Ery-CO3 were higher (145 mV $decade^{-1}$) as compared to Ery-PO4 (72 mV $decade^{-1}$), which was evaluated with the same conditions previously,[40] indicating an electrolyte-dependent mechanism. Therefore, the increase of $i_{max}$ for Ery-CO3 was most likely due to an increase in the number of active sites in the restructured catalyst with cycling rather than an increase in the efficiency of the active sites.

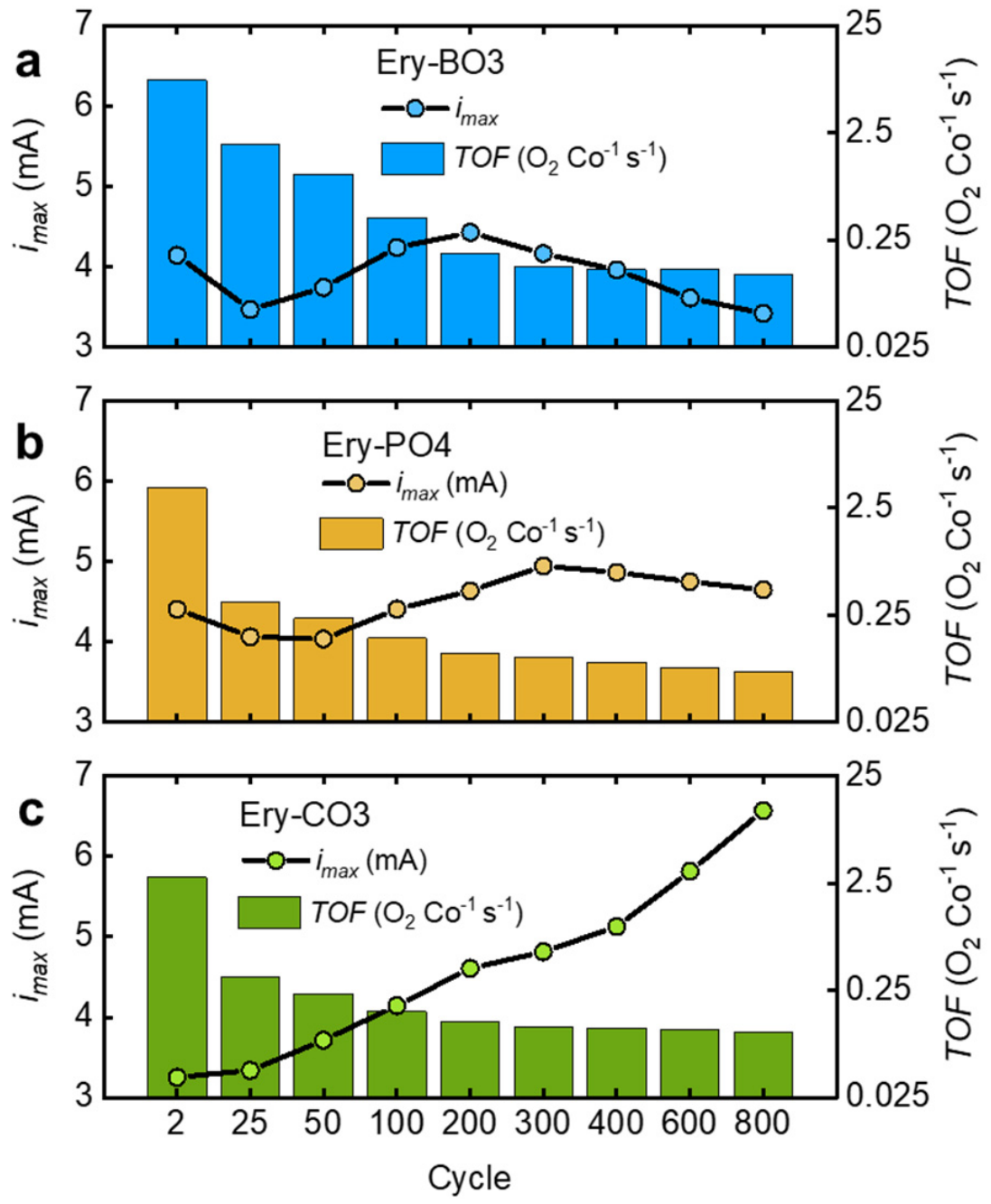

**Figure 6.** Relationship between catalytic current, $i_{max}$, (left y-axis, symbols) and TOF per redox-active site (right y-axis, bars) as a function of cycling in: **a.** Borate electrolyte, **b.** Phosphate electrolyte and **c.** Carbonate electrolyte. $i_{max}$ was evaluated at 2.1 V vs. RHE. TOF was estimated by the normalization of $i_{max}$ by ERC and the electrons transferred (4 electrons).

## 3. Summary and Conclusion

Ery was used to identify the requirements for catalyst activation by electrochemical restructuring, for which we defined the indicators of the redox active Co (ERC), loss of Ery crystal structure and anion exchange. All investigated electrolytes led to electrochemical restructuring to an amorphous layered cobalt (hydr)oxide to various degrees as demonstrated by EXAFS analysis. In Ery-BO3, all three restructuring indicators correlated with cycling. In Ery-PO4, only two indicators correlated: the loss of the Ery structure and arsenate, while the ERC increased monotonously. In the most interesting case of Ery-CO3, none of the indicators correlated over 800 cycles. This indicated that the kinetics of the restructuring processes, both structural and compositional, differ strongly in the selected electrolytes. Furthermore, the ERC correlates with the number of active sites, yet its continuous increase resulted in the desired increase in activity only for Ery-CO3. This highlights that the restructuring and activation are two separate processes, where activation has further requirements.

For activation, a large cluster size (i.e., high CPR) and high Co valence of the restructured surface (hydr)oxide were beneficial. An adequate cluster size has been reported to play an essential role in the mesostructure arrangement, which, together with intercalated electrolyte anions, impacts the electrical conductivity properties of the material.[27,29] We hypothesize that a larger cluster size also aids in keeping anions in the amorphous structure where they act a proton buffers[13,61] so that larger clusters can support higher currents. Finally, we found that the current increased with the Co oxidation state at fixed cluster size where an average oxidation state of 2.8 produced the highest activity. This is a typical oxidation state of a layered double hydroxide[31,59,62] and close to ex situ measurements of electrodeposited Co oxide.[48,49]

In summary, our work highlights the role of the electrolyte for electrochemical restructuring, where we found surprisingly different behavior of the restructured Ery. We expect that the simple method of anion exchange can be utilized in a wide range of catalyst materials, e.g. (oxy)phosphates,[12,13] fluorides,[63] oxychlorides,[23] multimetallic oxyhydroxides,[18,24,64,65] perovskites,[35] to control catalyst activation by restructuring with an optimal trinity of local order, transition metal valence and high number of active sites.

## 4. Materials and methods

### 4.1. Materials

$CoSO_4 \cdot 7H_2O$ (Merck, 99,5%), $Na_2HAsO_4 \cdot 7H_2O$ (Merck, 99,5%), Nafion™117 5% in etanol (Sigma-Aldrich) , FTO Pilkinton NSG TEC T15, $K_2HPO_4$ (Riedel-De Haën, 98%), $KH_2PO_4$ (Merck, > 99,5%) $H_3BO_3$ (Sigma-Aldrich > 99,5%), HCl (Merk, 32%) , $NaHCO_3$ (Merk, 99,5%), isopropanol (Labquimar, 99,9%), MiliQ water (>18 MΩ cm), $CaCl_2$ (Merk, p.a.).

### 4.2. Synthesis of Erythrite (Ery)

400 mL of a 7 mM $Na_2HAsO_4 \cdot 7H_2O$ solution was added dropwise to 800 mL of 5 mM $CoSO_4$ solution at 65 °C. The reddish solution of Co turned pink after several drops, and a precipitate started forming. The final suspension was stirred and heated (at 65 °C) for 72 h. The obtained solid was washed five times with deionized water and dried at room temperature and vacuum using $CaCl_2$ as a desiccant. The resulting product is denoted Ery herein.

### 4.3. Electrode preparation

5.00 mg of Ery were added to 1000 µL of isopropanol, then the suspension was sonicated for 2 h to get a higher dispersion. 80 µL of this suspension was slowly drop-coated on a 1 cm x 1 cm FTO surface glass. Once the isopropanol evaporated, 15 µL of 0.25 % Nafion were drop-coated three times on the electrode's surface, waiting 5 minutes between each addition.

### 4.4. Electrolyte preparation

All electrolytes were prepared at pH 7 (measured with a Daigger 5500 pH Meter) and 0.1 M concentration. Phosphate: 0.1 M $K_2HPO_4$ and 0.1 M $KH_2PO_4$ solutions were prepared and mixed in an adequate ratio to adjust to pH 7. Borate electrolyte: a 0.1 M $B_4Na_2O_7$ solution was prepared, and 0.1 M HCl solution was added dropwise to adjust to pH 7. Carbonate electrolyte: a 0.1 M $NaHCO_3$ solution was prepared, and 0.1 M HCl solution was added dropwise to adjust to pH 7. Arsenate electrolyte: a 0.1 M $Na_2HAsO_4 \cdot 7H_2O$ solution was prepared, and 0.1 M HCl solution was added dropwise to adjust to pH 7.

### 4.5. Electrochemical measurements

The experiments were carried out on FTO glass electrodes as working electrode in a single-compartment three-electrode electrochemical cell made of glass filled with about 50 mL solution of the electrolyte. We used a high-surface Pt mesh counter electrode and an Ag/AgCl (KCl saturated) reference electrode (separation of about 1 cm). The electrochemical experiments were performed at room temperature using a potentiostat (SP-300, BioLogic Science Instruments) controlled by the EC-Lab v11.01 software package. The typical electrolyte resistance (incl. the electrode) was about 65 Ω; iR compensation at 85% was dynamically applied. The solution remained unstirred during the experiments. All potentials were calculated and converted with respect to the reversible hydrogen electrode (RHE).

### 4.6. X-ray diffraction (XRD)

Diffractograms were collected using a Bruker D8 Göbel-Mirror for grazing incidence in PT006 and an energy-dispersive Sol-X detector. Cu-Anode (Kα1+2) was used as source of X-ray. Data were collected in the range of 2Θ=10 ° to 70 ° with increments of 0.02° and an equivalent time of 10 s per step. The measurements were carried out at the X-ray Core lab at Helmholtz-Zentrum Berlin. We used the (020) reflection of Ery (2Θ = 13°) normalized by the (200) reflection of FTO (2Θ = 38°) as an indicator of the crystalline Ery phase.

### 4.7. Scanning electron microscopy (SEM) and energy-dispersive X-ray spectroscopy (EDX)

Sample morphology was determined using a low vacuum scanning electron microscope, HITACHI S-3700N, with an acceleration voltage of 10 keV and detecting secondary electrons. EDX measurements were performed using an energy dispersive X-ray analysis probe, Oxford, and a Hitachi S-570 with an Aspe model Sirius 10/7.5 EDS. Normalization by the amount of Co guarantees that changes in As are due not to the dissolution of the material itself.

### 4.8. X-ray absorption spectroscopy (XAS)

Hard XAS spectra at Co-K edge were collected at the KMC-3 beamline[66] at Helmholtz-Zentrum Berlin. Spectra were recorded in fluorescence mode using a 13-element Silicon Drift Detector (SDD) from RaySpec. The used monochromator was a double-crystal Si(111), and the polarization of the beam was horizontal. Reference samples were prepared by dispersing a thin and homogeneous layer of the ground powder on Kapton tape. After removing the excess material, the tape was sealed, and the excess of Kapton was folded several times to get 1 cm x 1 cm windows. The energy was calibrated using a Co metal foil (fitted reference energy of 7709 eV in the first derivative spectrum) with an accuracy ±0.1 eV. Up to three scans of each sample were collected to k = 14 $Å^{-1}$.

All spectra were normalized by subtracting a straight line obtained by fitting the data before the K edge and division by a polynomial function obtained by fitting the data after the K edge, as illustrated elsewhere.[14] The Fourier transform (FT) of the extended X-ray absorption fine structure (EXAFS) was calculated between 15 and 800 eV above the Co-K edge ($E_0$=7709 eV). A cosine window covering 10 % on the left side and 10 % on the right side of the EXAFS spectra was used to suppress the sidelobes in the FTs. EXAFS simulations were performed using the software SimXLite.

After calculating the phase functions with the FEFF8-Lite[67] program (version 8.5.3, self-consistent field option activated), atomic coordinates of the FEFF input files were generated from the structure of Ery and other several reasonable structural models (**Figure S1**); the EXAFS phase functions did not depend strongly on the details of the used model. An amplitude reduction factor ($S_0^2$) of 0.85 was used, which is typical for Co oxides.[40] The data range used in the simulation was 34 to 747 eV (3.0 to 14.0 $Å^{-1}$) above the Co-K edge ($E_0$=7709 eV). The EXAFS simulations were optimized by minimizing the error sum obtained by summation of the squared deviations between measured and simulated values (least-squares fit). The fit was performed using the Levenberg–Marquardt method with numerical derivatives.

**Supporting Information**
Supporting Information is available from the Wiley Online Library or from the author.

## 5. Acknowledgments

We thank Dr. Dulce Morales, Denis Antipin, and Joaquin Morales for helping in XAS data collection, and David Sánchez and Gabriela Fernández for helping in electrochemical measurements. We acknowledge Dr. Michael Haumann and Dr. Ivo Zizak for the support at the beamline. Dr. Katharina Klingan and Prof. Dr. Holger Dau for fruitful discussion. Dr. Petko

Chernev for permission to use his software SimXLite. The XAS experiments were financially supported by funds allocated to Prof. Holger Dau (Freie Univ. Berlin) by the Bundesministerium für Bildung und Forschung (BMBF, 05K19KE1, OPERANDO-XAS) and by the Deutsche Forschungsgemeinschaft (DFG, German Research Foundation) under Germany´s Excellence Strategy – EXC 2008 – 390540038 – UniSysCat. We also thank HZB X-ray CoreLab for training and advising in XRD. This project has received funding from Posgrado en Química and Vicerrectoría de Investigación (UCR), CONICIT-MICIT (Costa Rica), and the European Research Council (ERC) under the European Union's Horizon 2020 research and innovation programme under grant agreement No 804092.

## 6. Conflict of interest

The authors declare no conflict of interest.

# Supporting Information

**Requirements for beneficial electrochemical reconstruction: A model study on a cobalt oxide in selected electrolytes**

*Javier Villalobos, Diego González-Flores*, Roberto Urcuyo, Mavis L. Montero, Götz Schuck, Paul Beyer, Marcel Risch**

## Methods

### *Fourier transform infrared spectroscopy (FTIR)*

Spectra were collected on the electrodes containing the material after a selected number of cycles in a Perkin Elmer Frontier FTIR-ATR mode using a diamond crystal. Measurements were performed using 64 scans between 600 and 4000 $cm^{-1}$.

### *Raman spectroscopy*

Spectra were acquired on a WITec Alpha 300R Raman microprobe system, with 750 nm excitation at a power of 20 mW and an integration time of 2 s with 20 accumulations. Measurements were replicated on three different areas of samples.

**Figures**

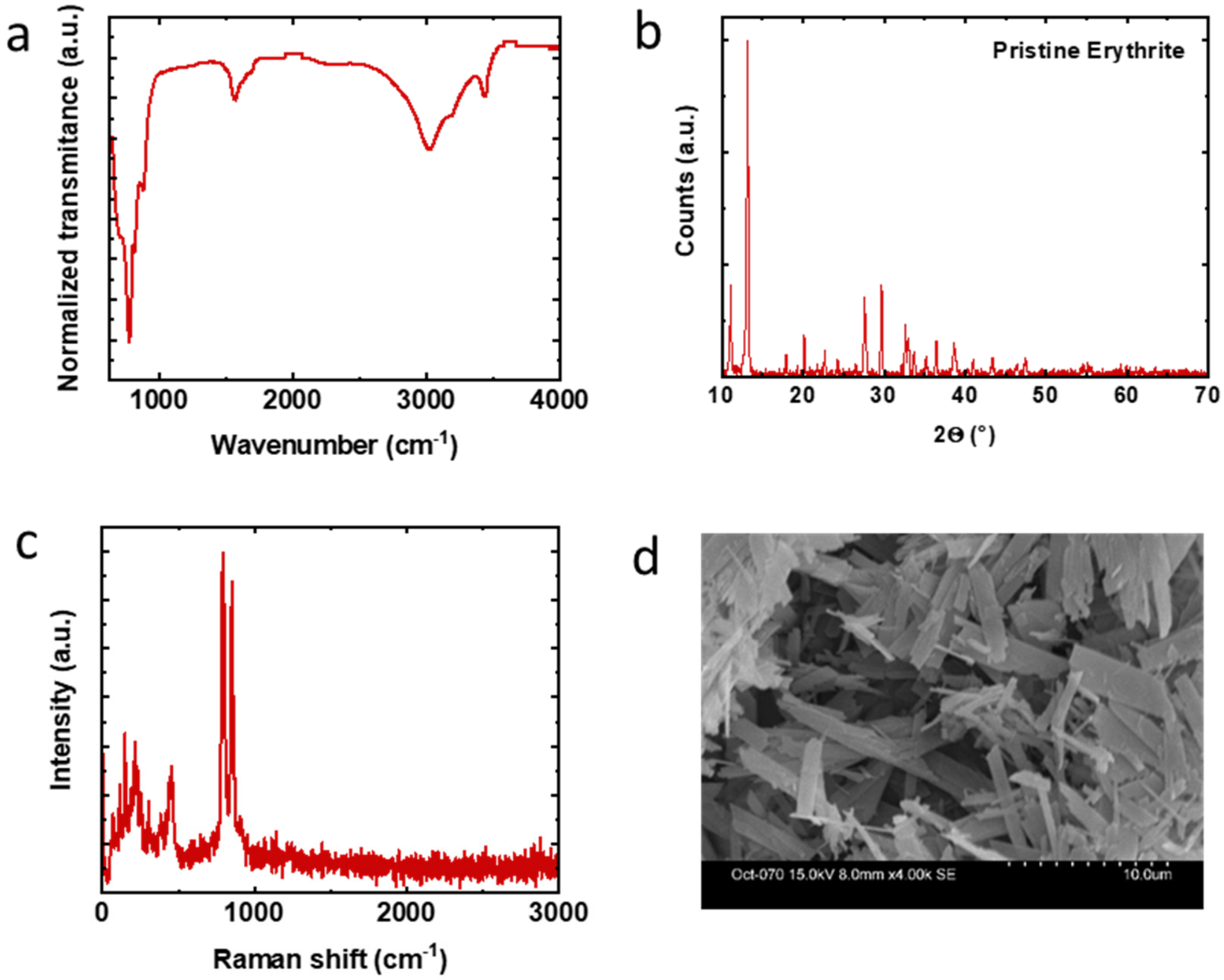

**Figure S1.** Characterization of Ery: a. ATR-FTIR spectrum, b. Powder X-ray diffractogram of Ery. The peak around 13 ° confirms the presence of the vivianite face, c. Raman spectrum, d. SEM images of some crystals.

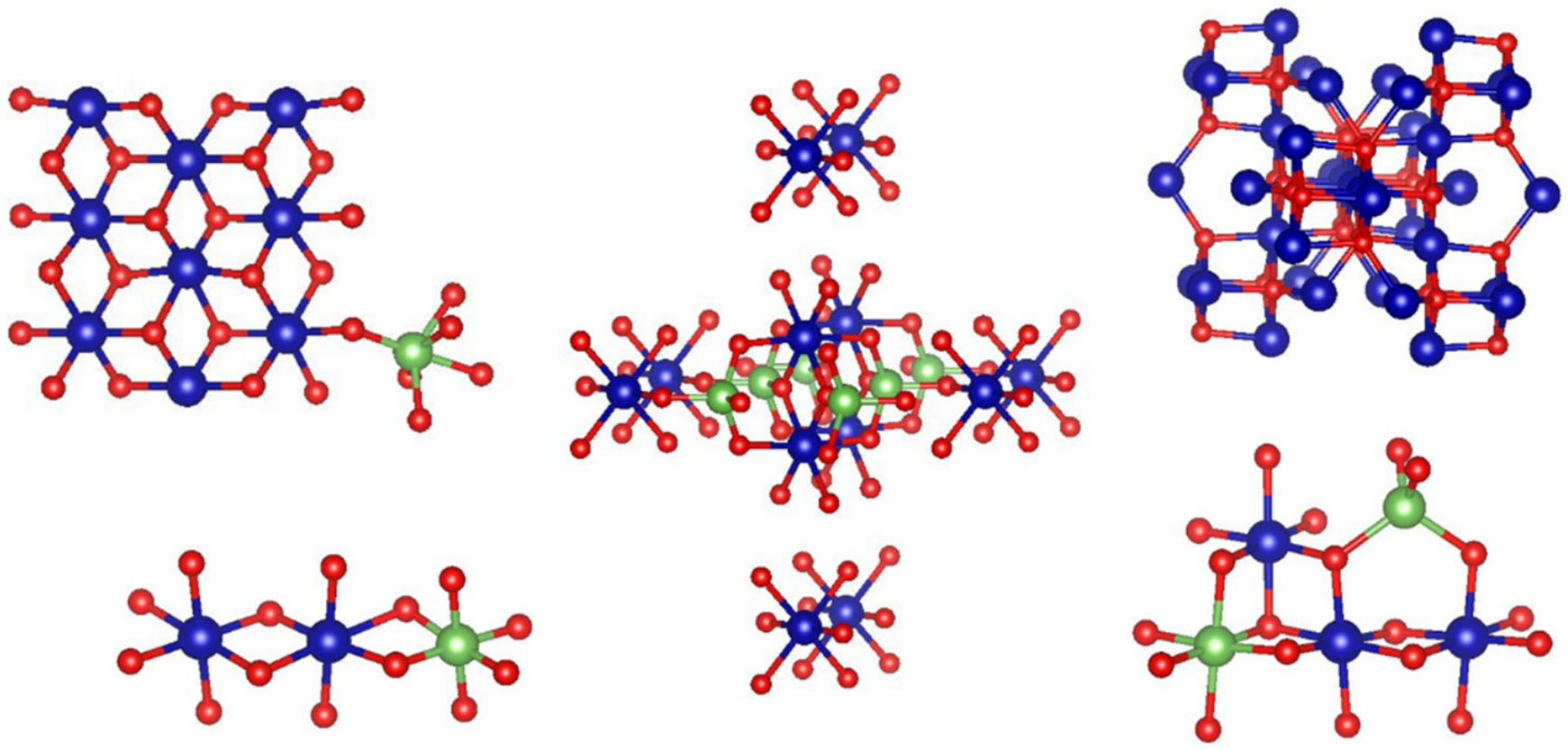

**Figure S2.** Structural models for the estimation of reasonable phase functions for EXAFS simulations. Blue dots represent Co atoms, red dots oxygen atoms, and green dots arsenic atoms.

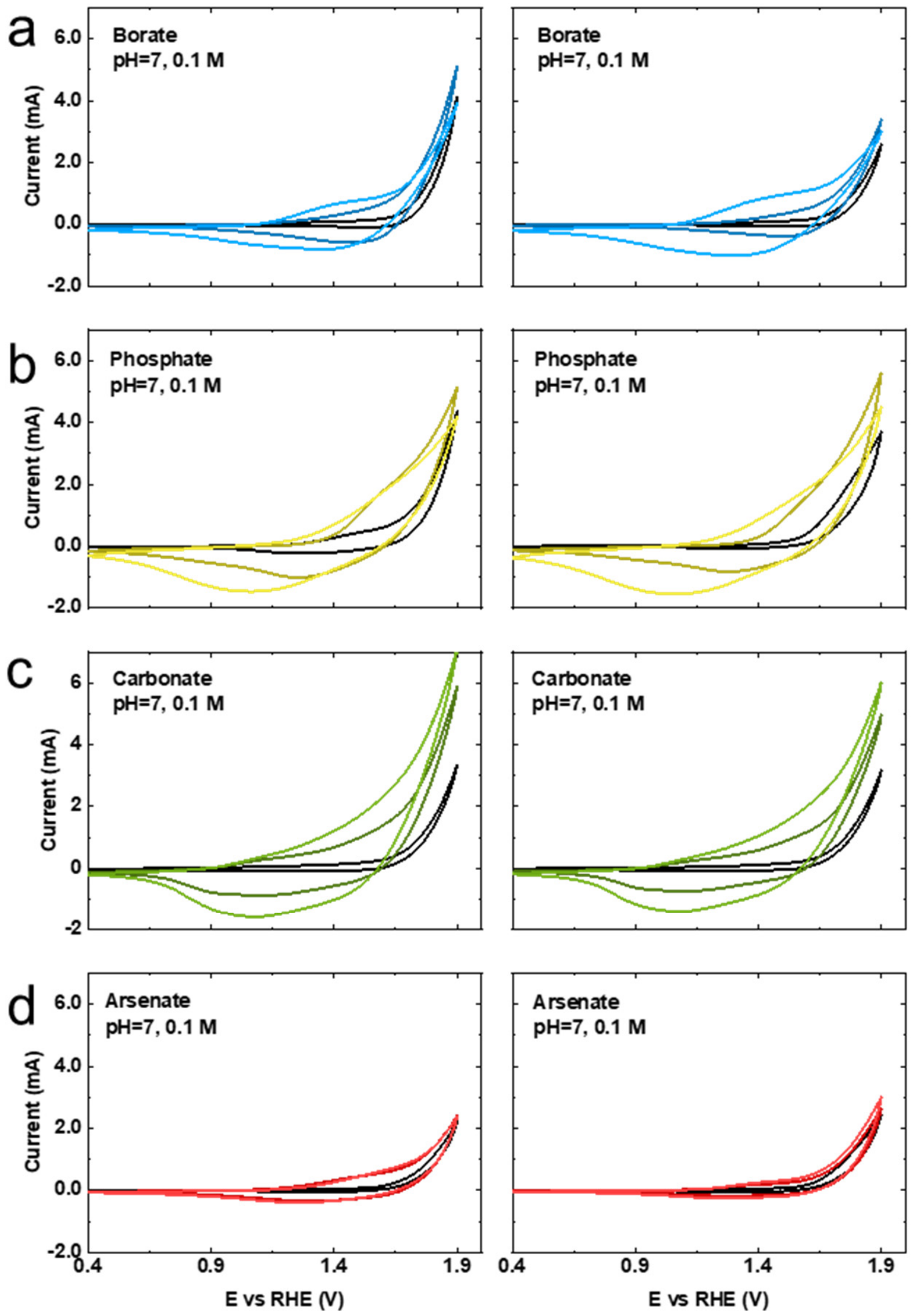

**Figure S3.** Series of CV performed on an Erythrite-deposited FTO glass (samples set b and c) in different electrolytes at pH 7 and a concentration of 0.1 M: **a.** Borate, **b.** Phosphate, **c.** Carbonate and **d.** Arsenate. 800 cycles were performed with a sweep speed of 100 mV $s^{-1}$ and 85 % iR compensation. Potential is represented as E vs. RHE.

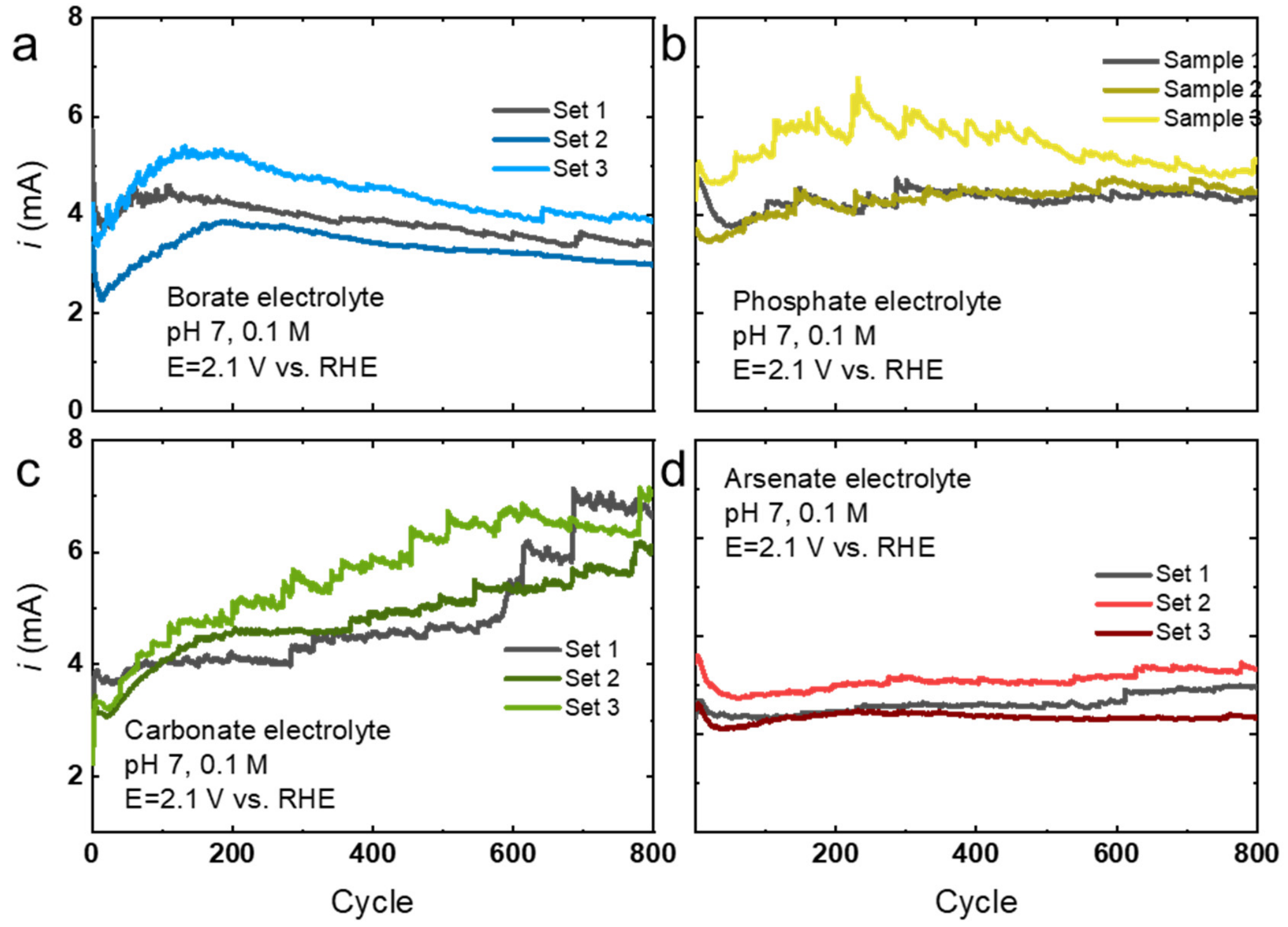

**Figure S4.** Maximum current ($i_{max}$) as a function of cycling of Ery-deposited FTO in: **a.** borate, **b.** phosphate, **c.** carbonate and **d.** arsenate electrolytes. All electrolytes at pH 7 and 0.1 M concentration. The three lines represent three different samples. $i_{max}$ were obtained at E=2.1 V vs. RHE.

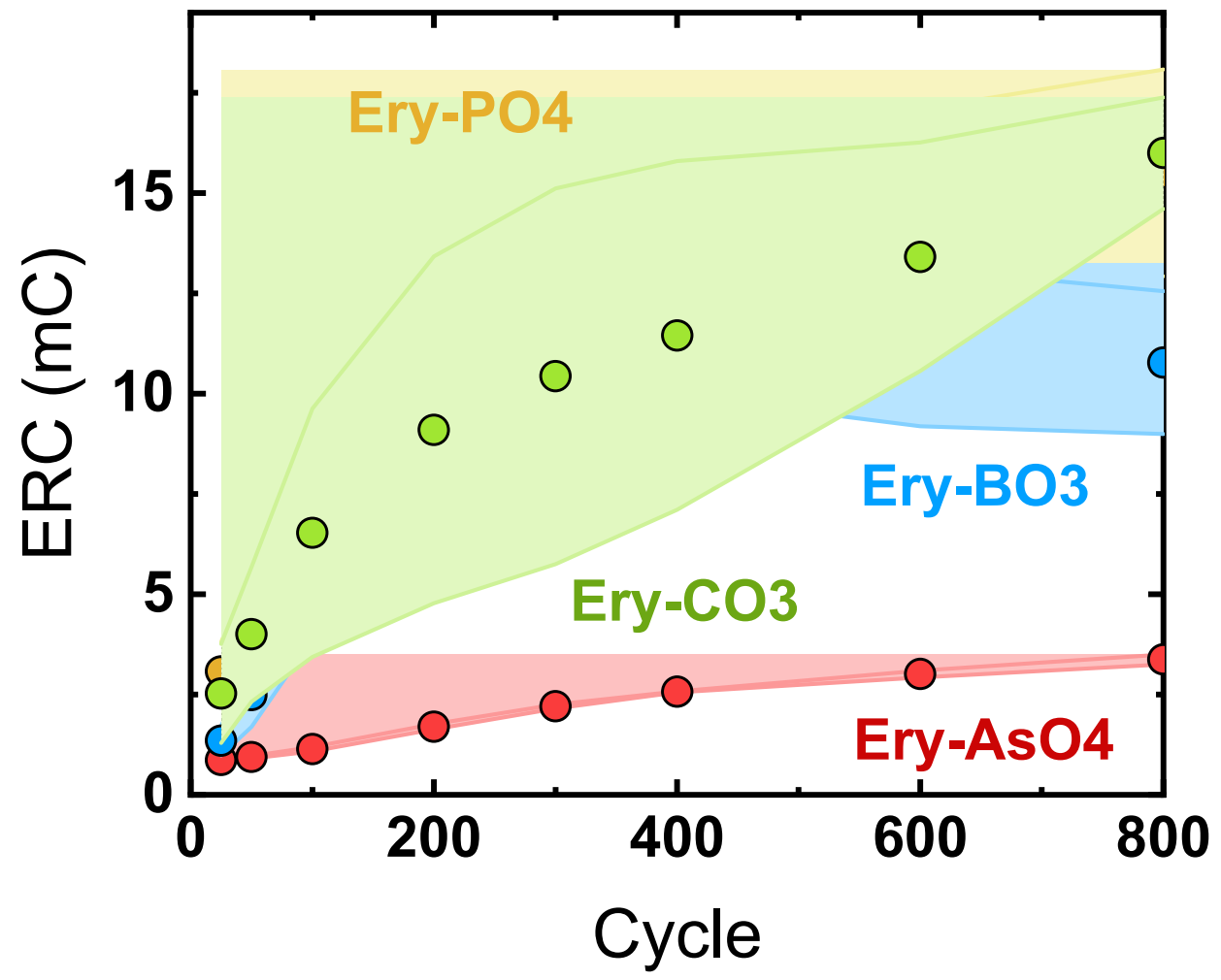

**Figure S5.** ERC as a function of cycling of Ery-deposited FTO in borate (blue), phosphate (yellow), carbonate (green) and arsenate (red) electrolytes. All electrolytes at pH 7 and 0.1 M concentration. The light-colored area represents the standard deviation of three values. ERC values were estimated by the integration of the reductive currents.

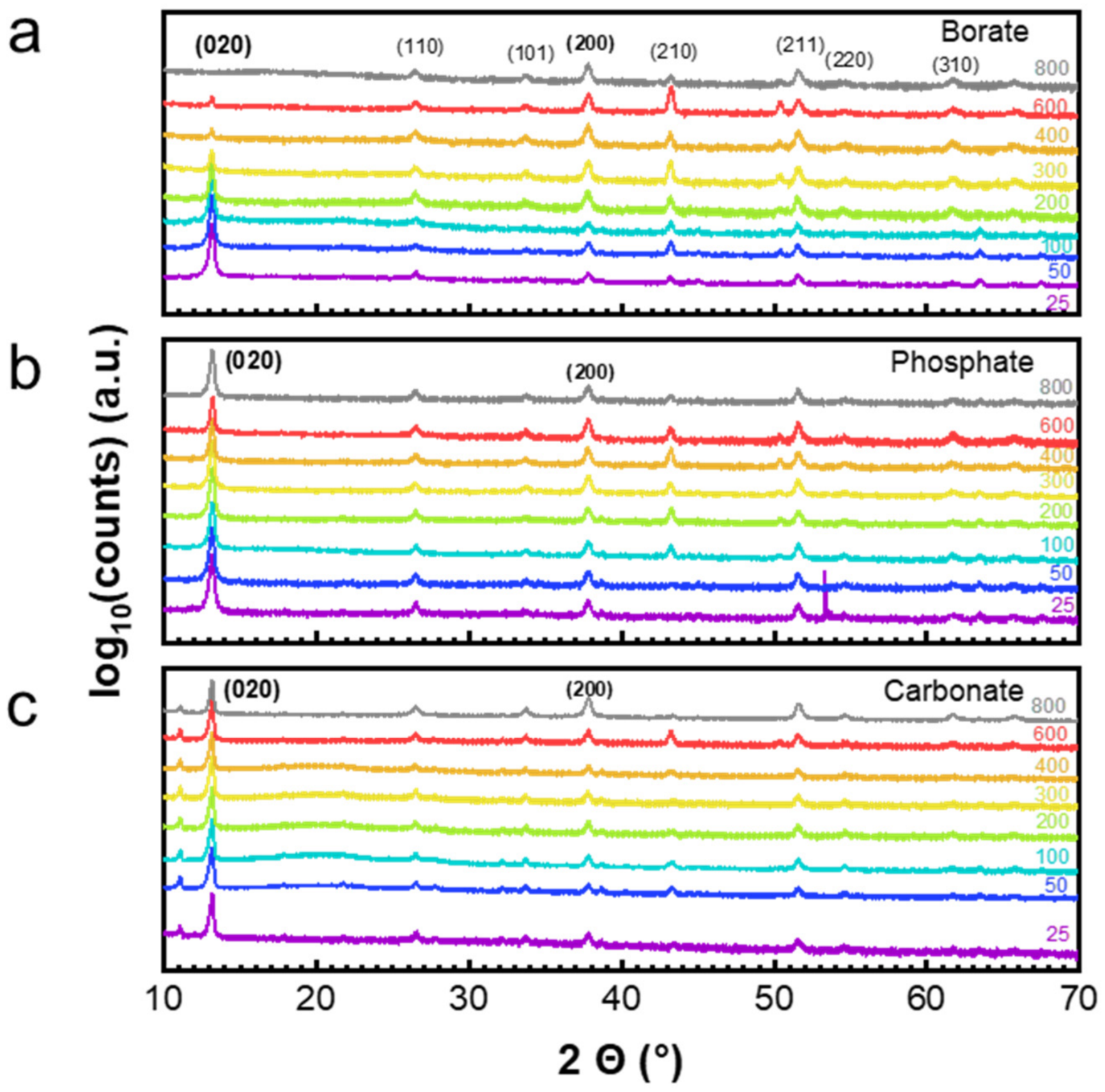

**Figure S6.** XRD patterns of Ery-deposited FTO glass after a selected number of cycles in: **a.** pH 7 0.1 M borate electrolyte, **b.** pH 7 0.1 M phosphate electrolyte and **c.** pH 7 0.1 M carbonate electrolyte.

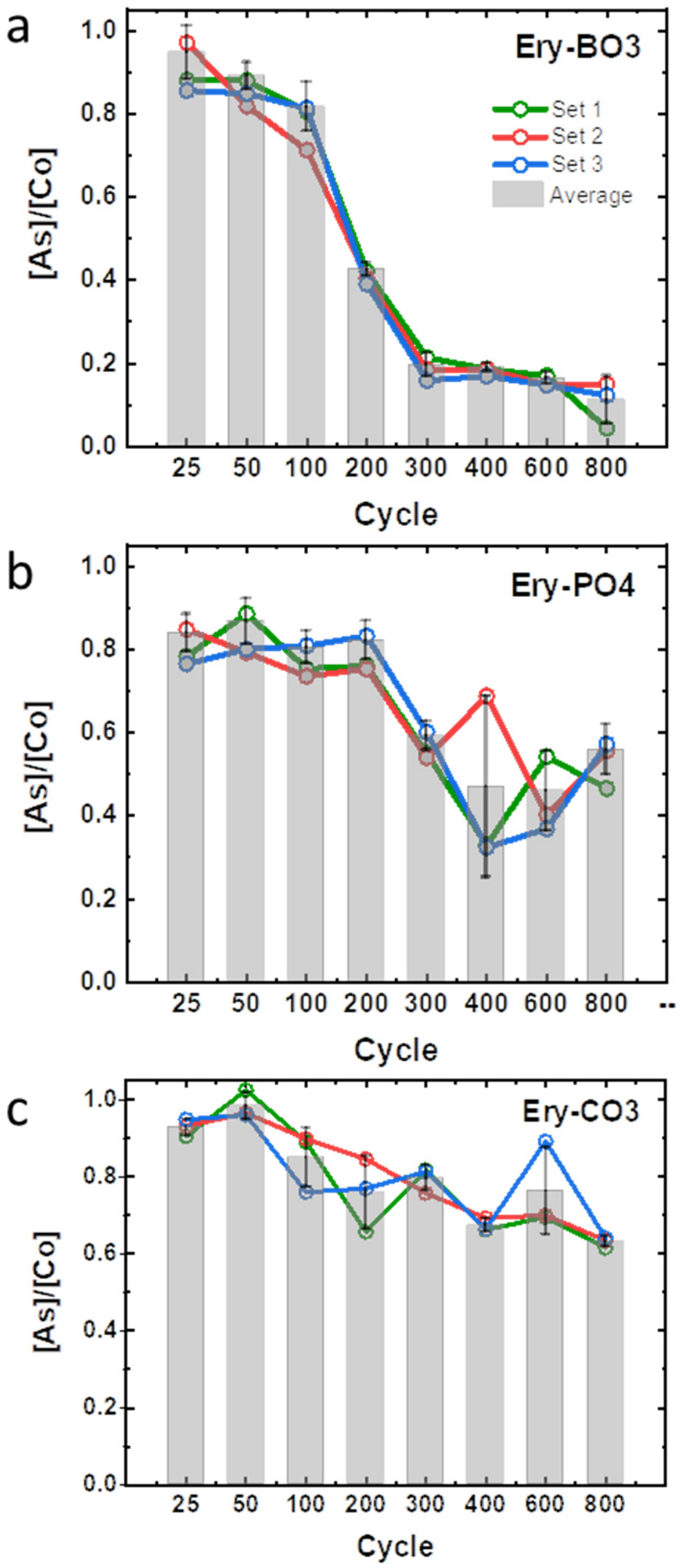

**Figure S7.** As/Co ratio by EDX as a function of cycling of Ery-deposited FTO glass in: **a.** Borate, **b.** Phosphate and **c.** Carbonate electrolyte. Three set of samples sets are represented with lines and the average is represented with bars (including standard deviation).

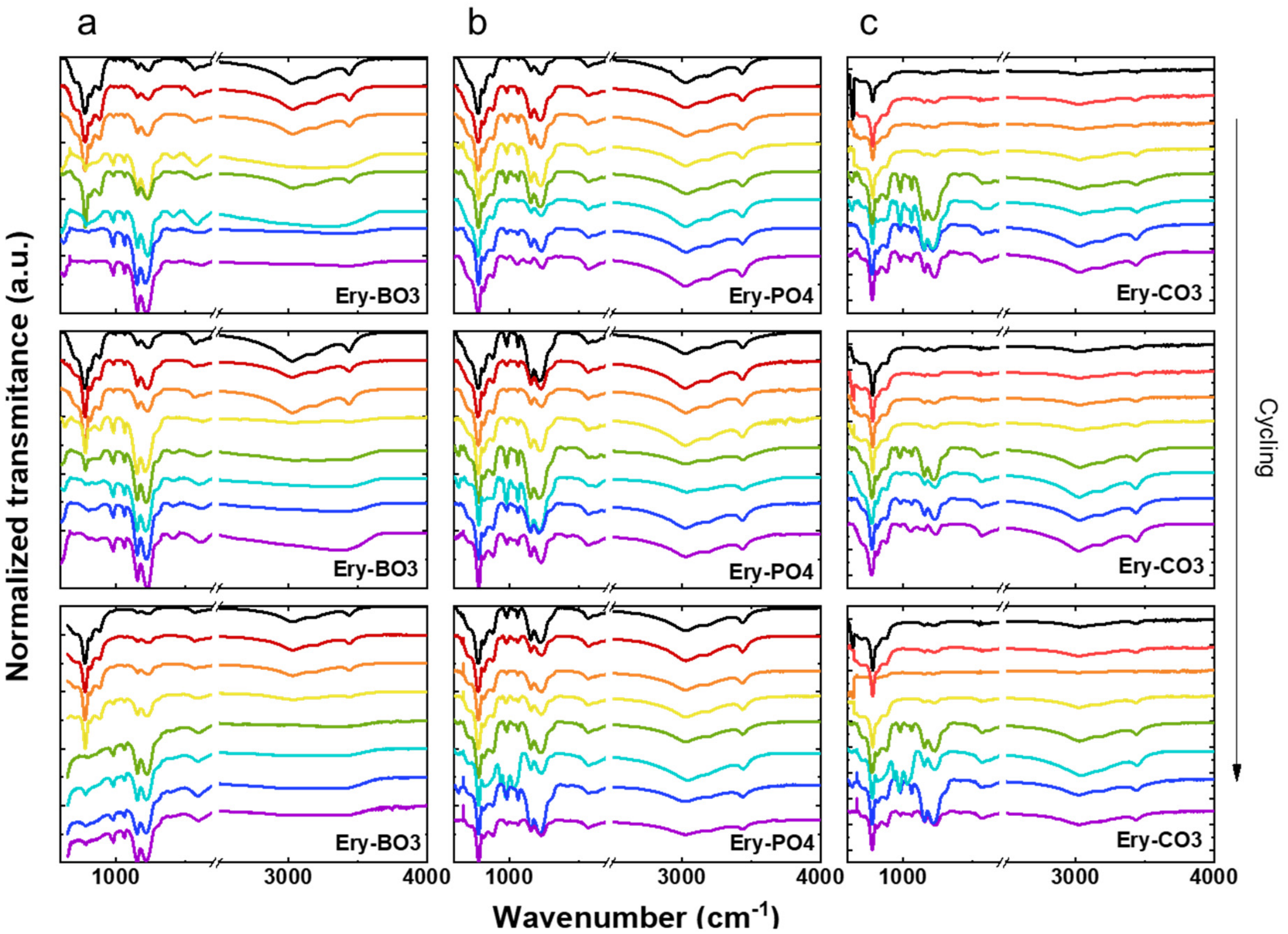

**Figure S8.** ATR-FTIR spectra of Ery-deposited FTO glass after a selected number of cycles (25, 50, 100, 200, 300, 400, 600, 800; descending) in pH 7 0.1 M: **a.** Borate, **b.** Phosphate and **c.** Carbonate electrolytes. The arsenate anions loss is tracked by the decrease at the 780 $cm^{-1}$ signal.

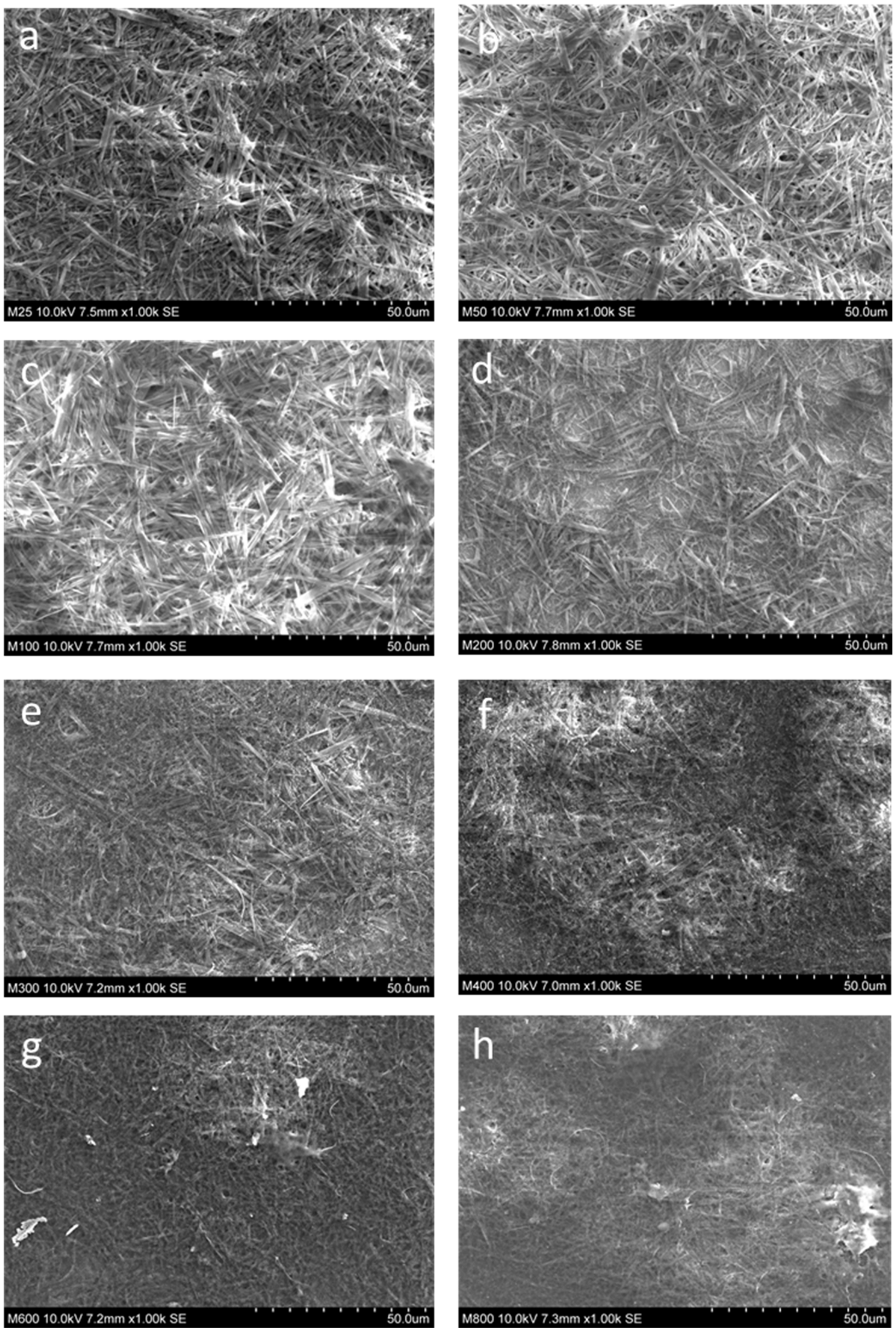

**Figure S9.** SEM images of Ery-deposited FTO glass in 0.1 M pH 7 borate electrolyte after a selected number of cycles: a. 25, b. 50, c. 100, d. 200, e. 300, f. 400, g. 600 and h. 800.

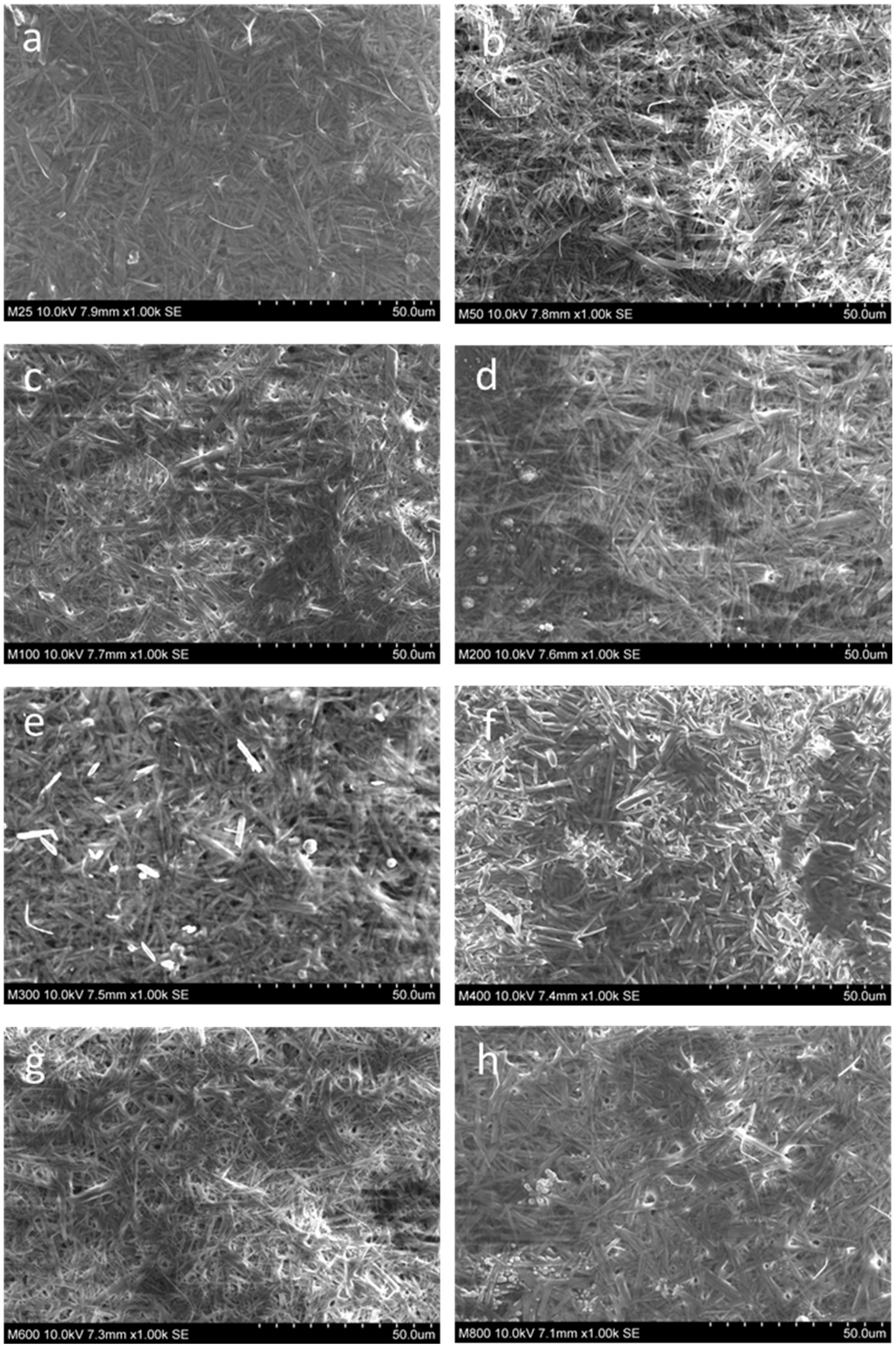

**Figure S10.** SEM images of Ery-deposited FTO glass in 0.1 M pH 7 phosphate electrolyte after a selected number of cycles: a. 25, b. 50, c. 100, d. 200, e. 300, f. 400, g. 600 and h. 800.

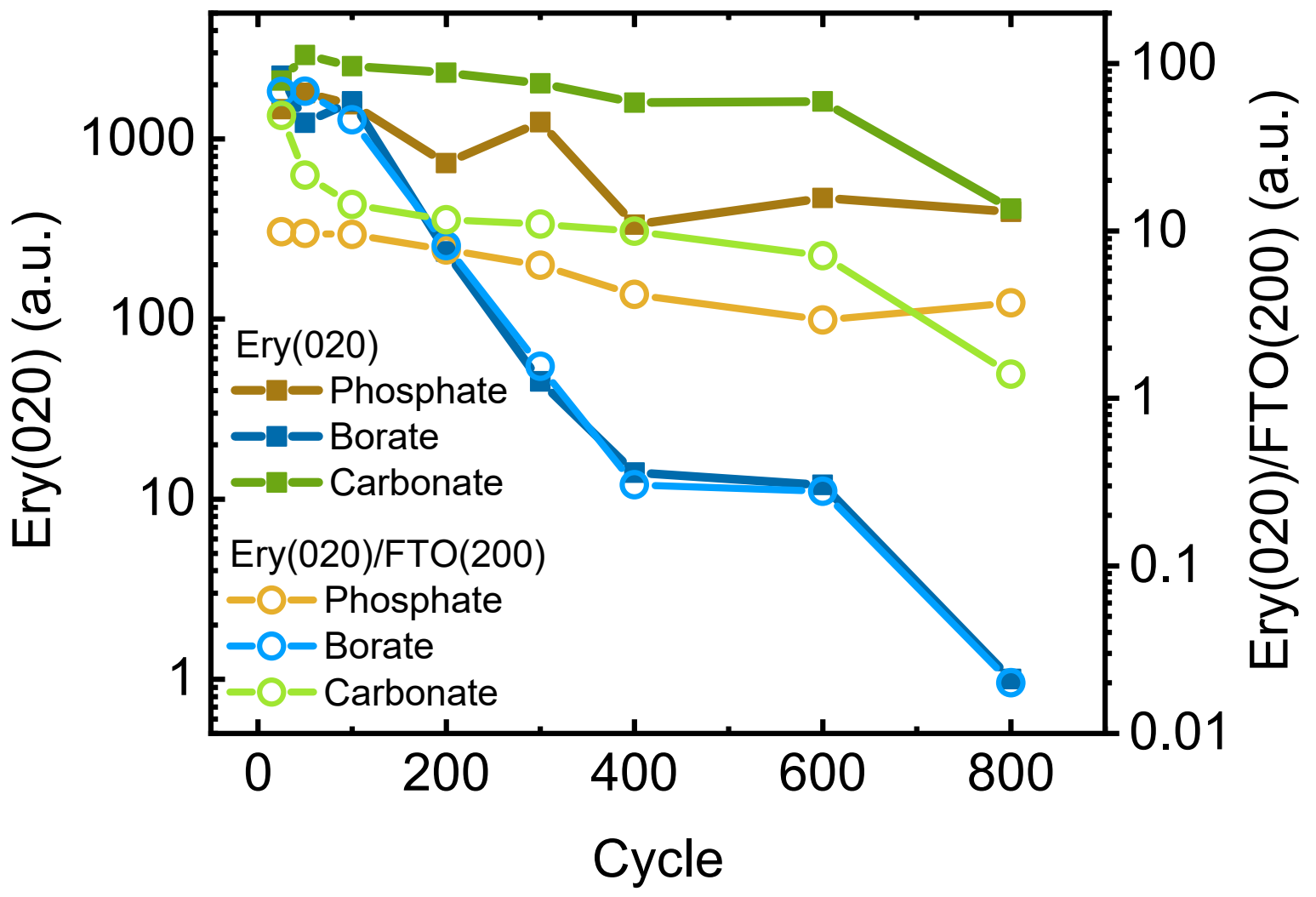

**Figure S11.** XRD integration peaks as a function of cycles for Ery-covered FTO substrates in borate and phosphate electrolyte. Ery(020) peak (left y-axis) and its normalization by the integration of FTO(200) (right y-axis).

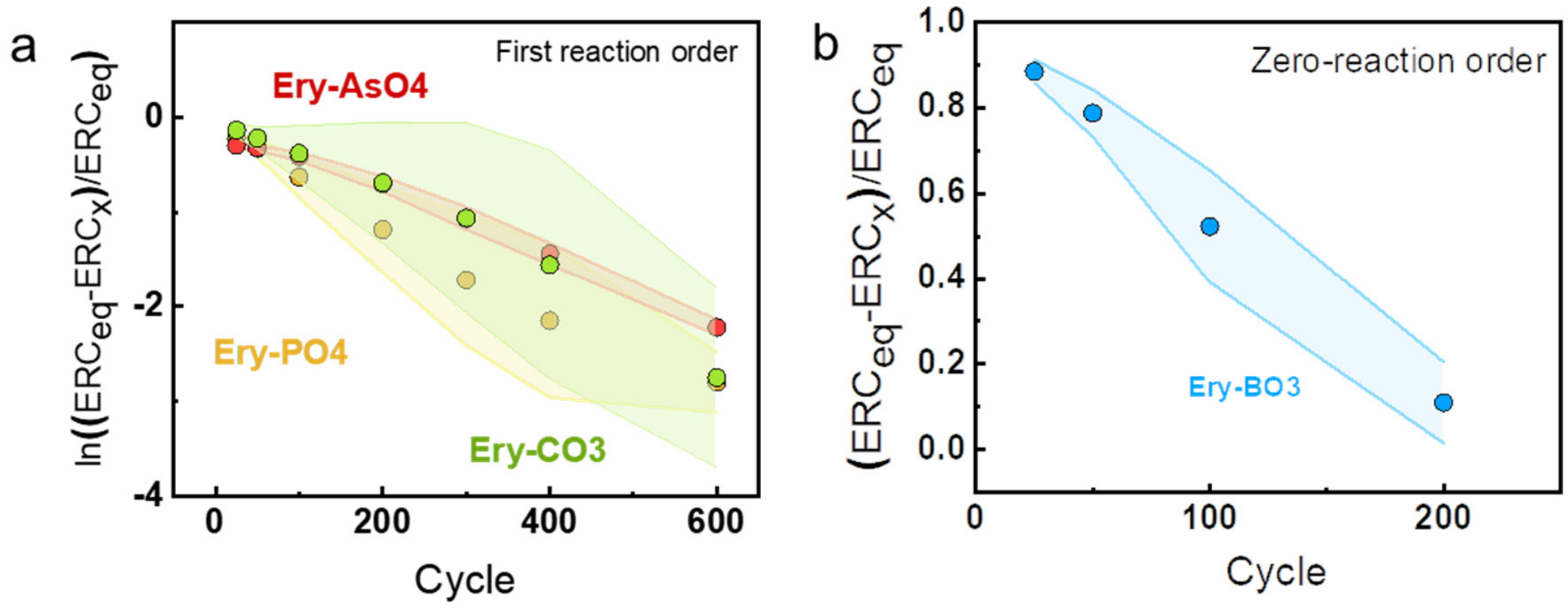

**Figure S12.** ERC reaction order fitting as a function of cycling of Ery in pH 7, 0.1 M borate (blue), phosphate (yellow), carbonate (green) and arsenate (red) electrolytes, **a.** Fitting equation: $\frac{ERC_{end}-ERC_i}{ERC_{end}}$ for zero-order reaction and **b.** Fitting equation: $ln\left(\frac{ERC_{end}-ERC_i}{ERC_{end}}\right)$ for first-order reaction. $ERC_i$ represents ERC's value in cycle $i$, and $ERC_{end}$ represents ERC's value at the final cycle (equilibrium). The light-colored area represents the standard deviation of three values.

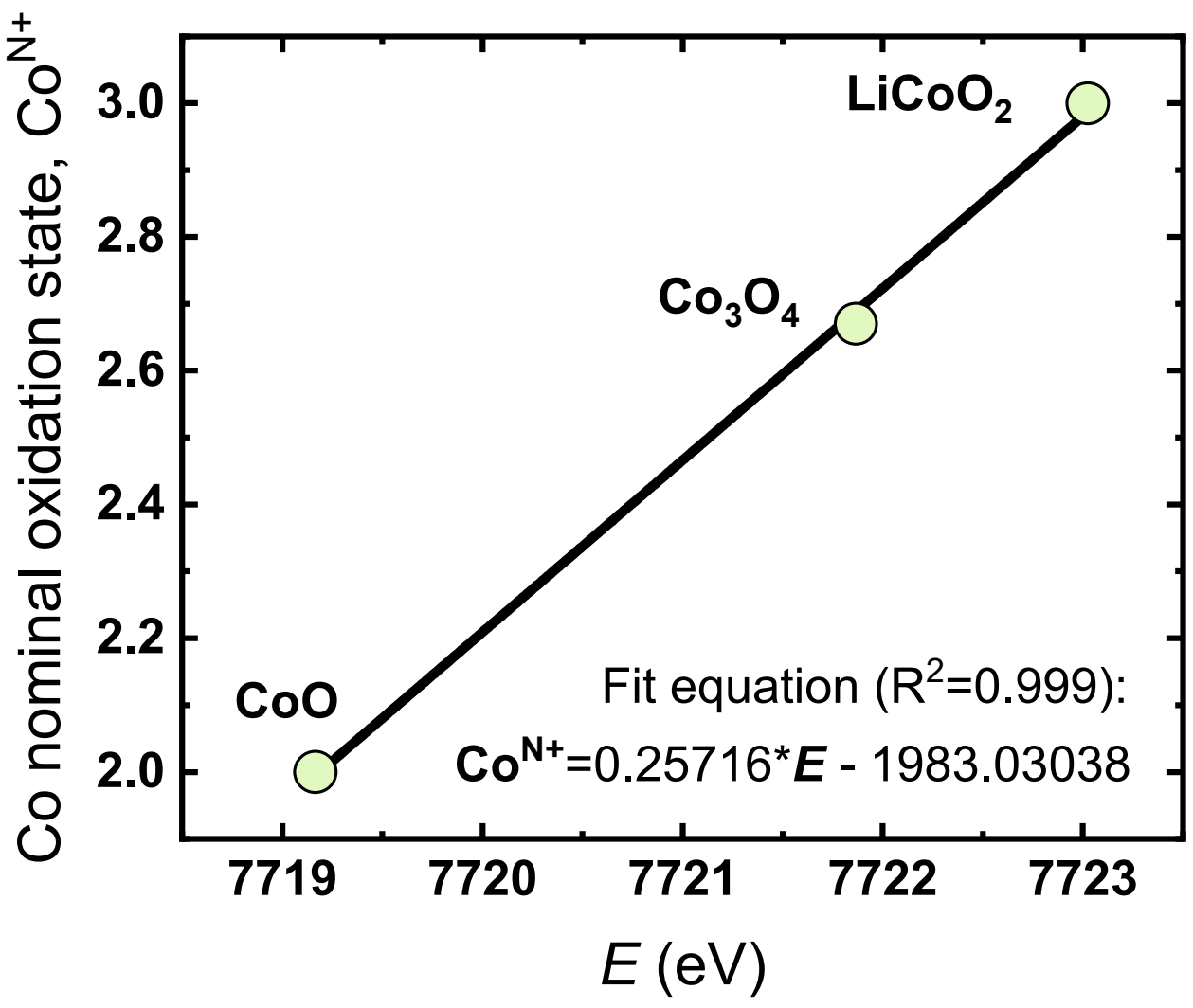

**Figure S13.** Co nominal oxidation state of Co oxides as a function of the Co-K edge E. The fit equation is shown in the graph. $Co^{(+2)}O$, $Co^{(+2.67)}{}_3O_4$ and $LiCo^{(+3)}O_2$ were used as references. The energy of the edge jump was estimated by the integral method.[1] The calculated nominal oxidation state for each sample is shown in **Table S**3.

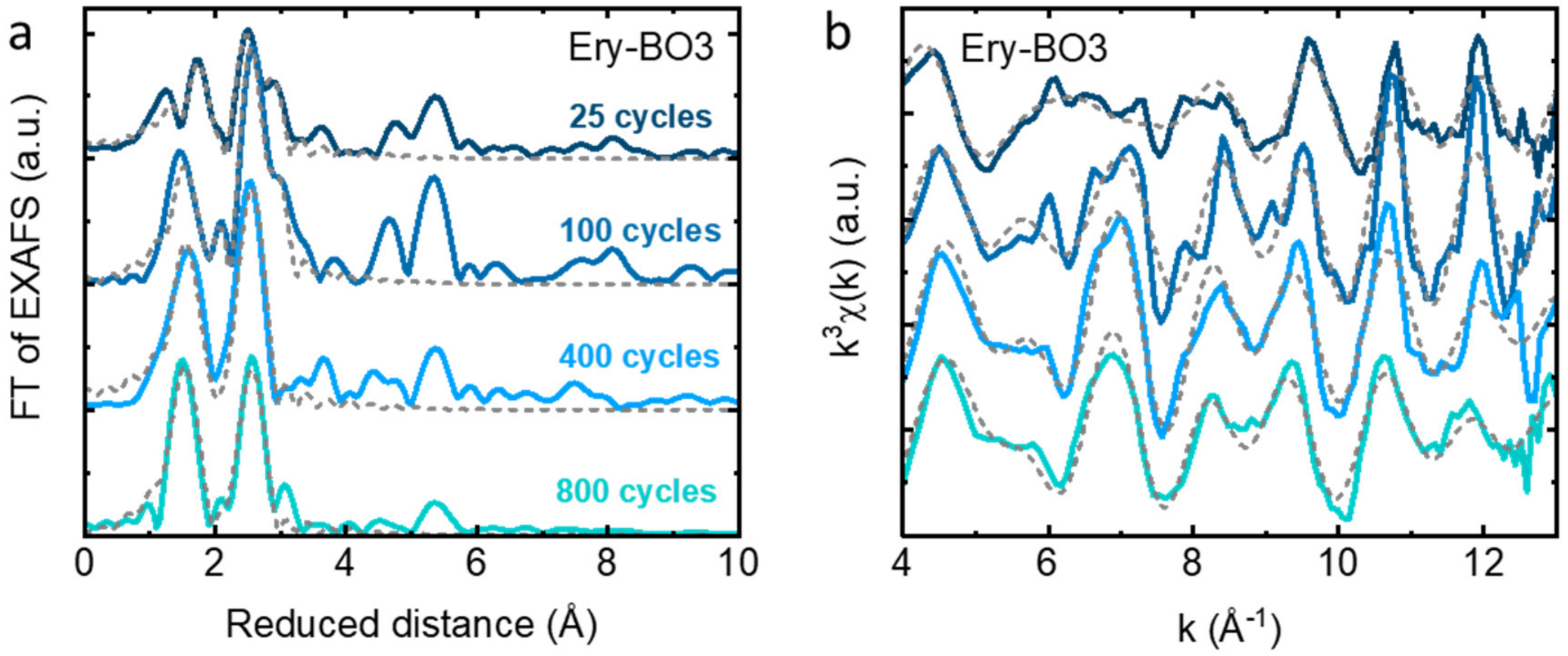

**Figure 14. a.** Fourier-transform EXAFS spectra for Co K-edge collected Ery-BO3 after a selected number of cycles. The corresponding dashed lines are results from EXAFS simulations (see **Table S4** for parameters). The reduced distance is by about 0.3 Å shorter than the precise distance obtained by EXAFS simulations. **b.** $k^3$-weighted EXAFS spectra of Ery-BO3 after a selected number of cycles recorded at the Co-K edge. The solid lines represent the measurements. The dashed lines represent the respective EXAFS simulations.

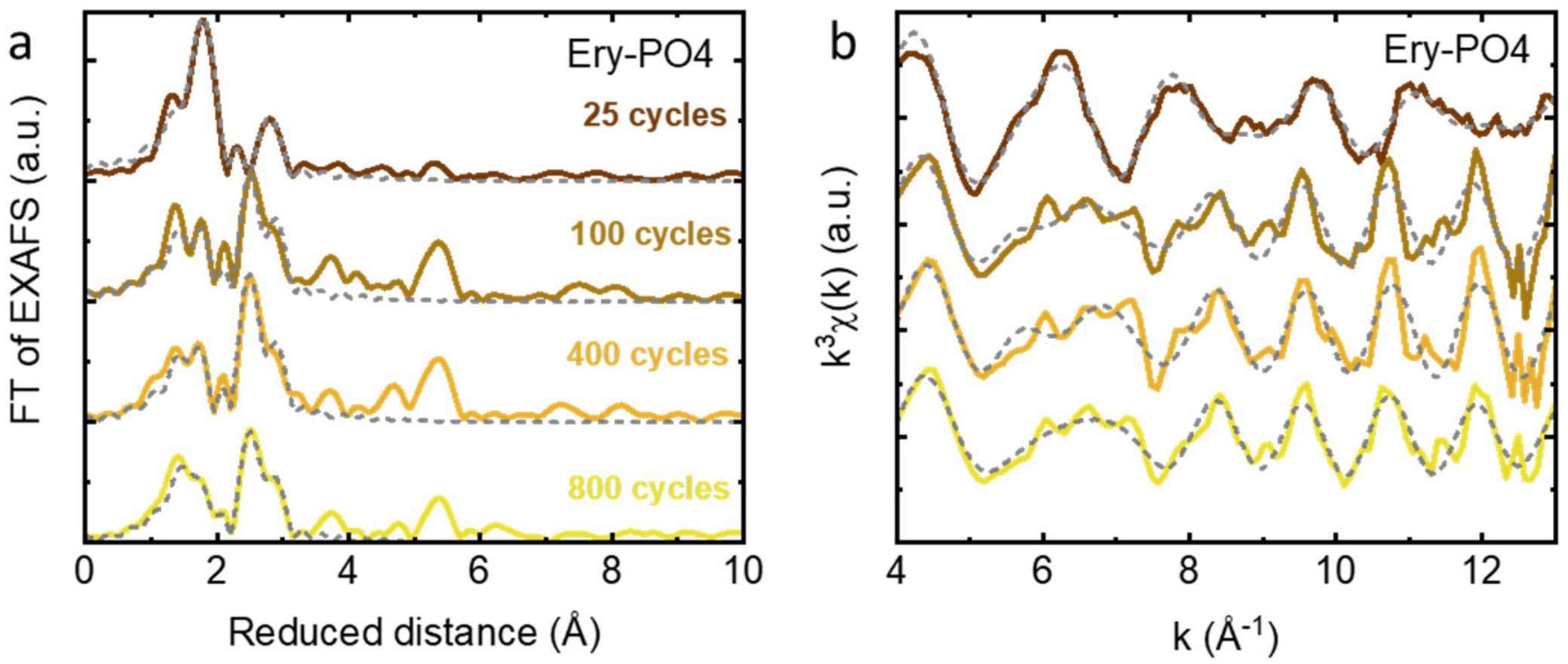

**Figure 15. a.** Fourier-transform EXAFS spectra for Co K-edge collected Ery-PO4 after a selected number of cycles. The corresponding dashed lines are results from EXAFS simulations (see **Table S4** for parameters). The reduced distance is by about 0.3 Å shorter than the precise distance obtained by EXAFS simulations. **b.** $k^3$-weighted EXAFS spectra of Ery-PO4 after a selected number of cycles recorded at the Co-K edge. The solid lines represent the measurements. The dashed lines represent the respective EXAFS simulations.

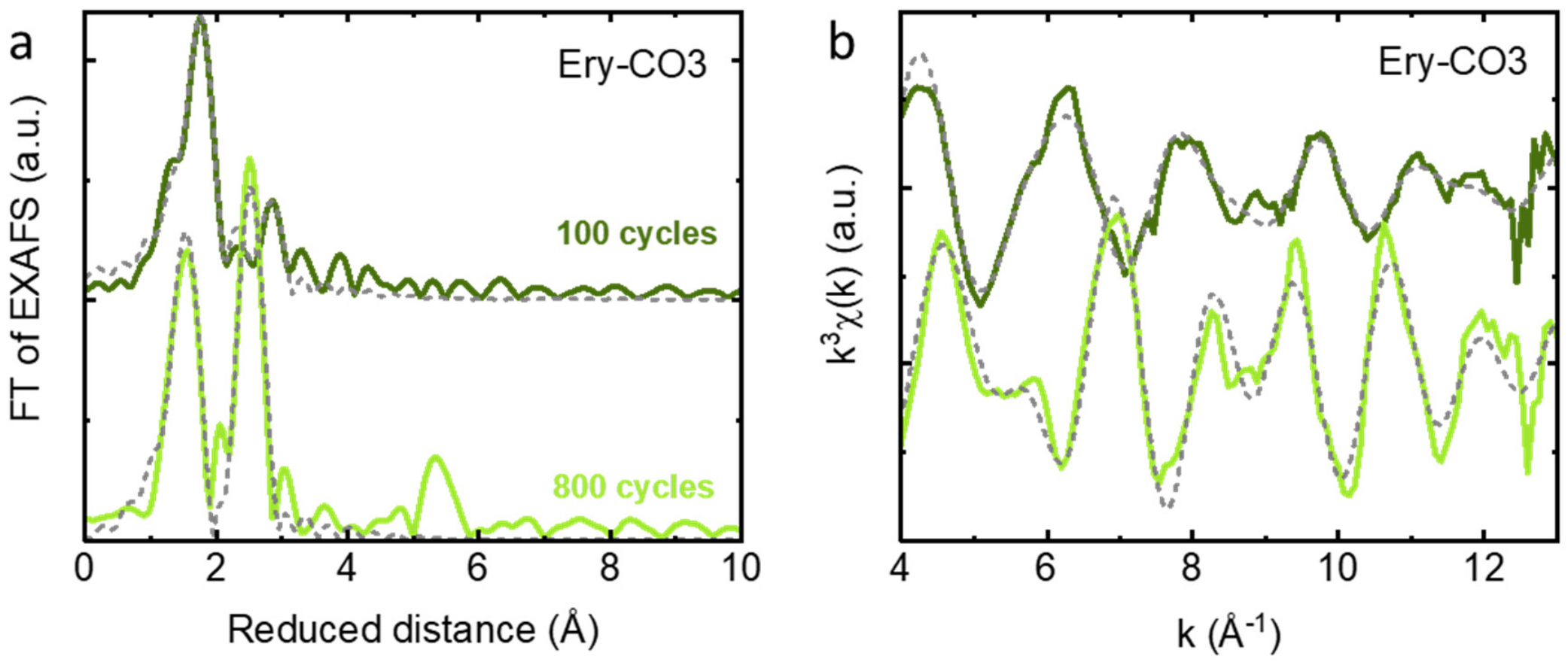

**Figure 16. a.** Fourier-transform EXAFS spectra for Co K-edge collected Ery-CO3 after a selected number of cycles. The corresponding dashed lines are results from EXAFS simulations (see **Table S4** for parameters). The reduced distance is by about 0.3 Å shorter than the precise distance obtained by EXAFS simulations. **b.** $k^3$-weighted EXAFS spectra of Ery-CO3 after a selected number of cycles recorded at the Co-K edge. The solid lines represent the measurements. The dashed lines represent the respective EXAFS simulations.

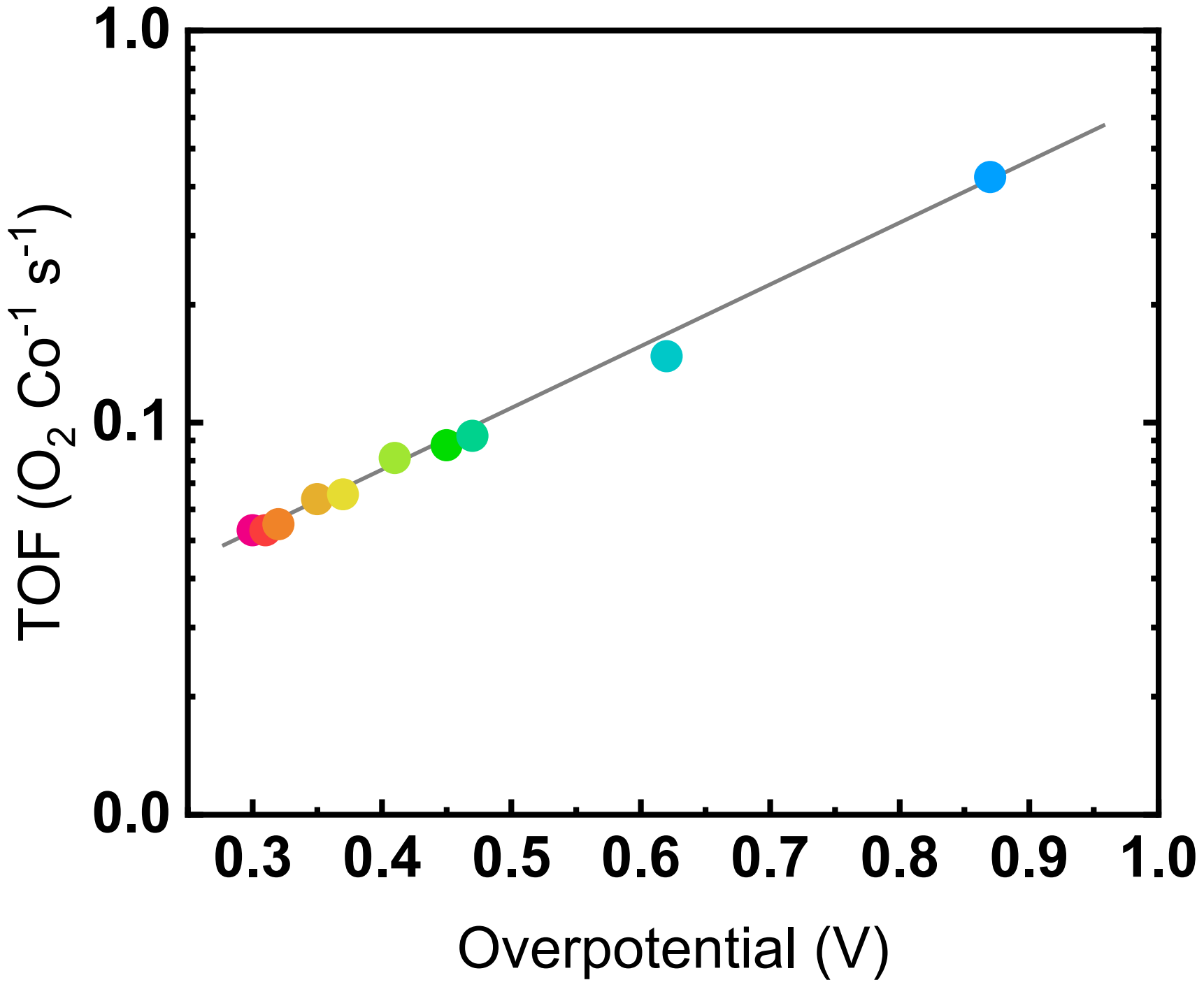

**Figure S17.** Turnover frequency (TOF) as a function of the overpotential of Ery-CO3 after 800 cycles. The values are estimated from the current and ERC shown in Figure 2 of the main text. The line was added to guide the eye.

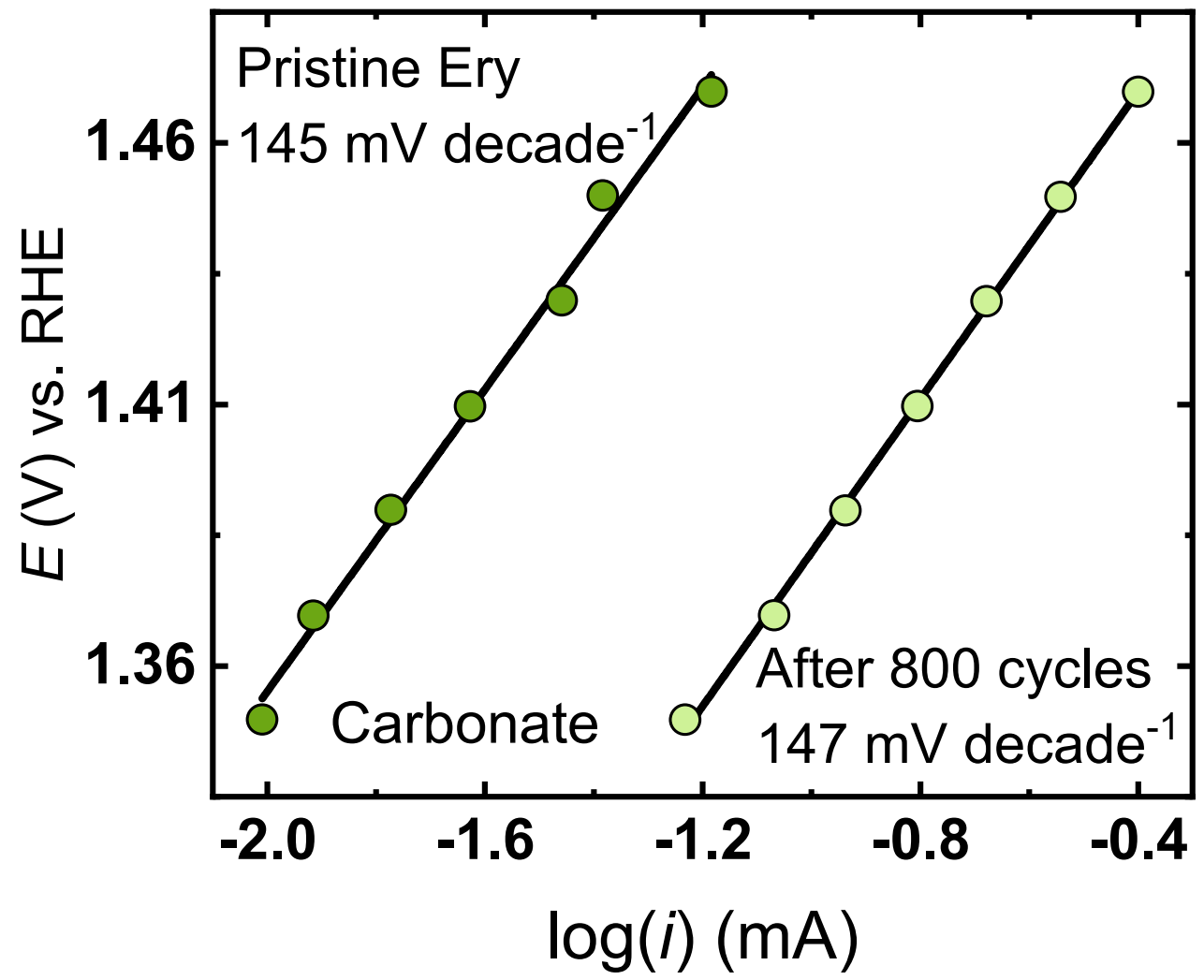

**Figure S18.** Tafel slope before and after reconstruction of Ery in carbonate pH 7 and 0.1 M concentration. The lines represent the linear fit of E-iR as function of log(i), the slope represents the Tafel slope. Tafel plots were collected with 50 mV steps, each potential was applied for 5 min. No iR compensation was applied. The potential was iR-corrected after the measurements.

## Tables

**Table S1.** General electrochemical protocol for data collection.

| **Step** | **Conditions** |
|---|---|
| **1. Cleaning** | FTO substrates were washed with soap to remove any grease on the surface and then, sonicated in isopropanol for 30 minutes to remove the soap and any dust. Lastly, they were sonicated for 30 minutes in deionized water. The substrates were dried at room conditions.<br>The glass cells were cleaned in a diluted solution of $HNO_3$ and then in deionized water. |
| **2. Reference electrode calibration** | The Ag/AgCl 3 M KCl reference electrode was calibrated against another Ag/AgCl 3 M KCl, which is only used with such purpose. |
| **3. Dynamic iR compensation** | Potential: 0.4 V vs. Ag/AgCl 3 M KCl<br>Frequency: 10 KHz<br>Amplitude: 10 mV<br>Waiting time between measurements: 0.1 s<br>Average from 4 measurements<br>85 % of compensation<br>Current range: 10 mA<br>Bandwith 7 fast |
| **4. Cyclic voltammetry** | Potential window: 0 V – 1.5 V vs. Ag/AgCl 3 M KCl.<br>Scan rate: 100 mV $s^{-1}$<br>Cycles: 25-800<br>iR dymanic compensation<br>Measured current over the last 50 % of the step duration<br>Recorded current averaged over 10 voltage steps<br>Current range: 10 mA<br>Bandwith 7 fast |
| **5. Sample rinsing** | After the electrochemical experiments the samples are soaked in deionized water during 5 minutes to remove the excess of electrolyte. |

**Table S2.** Fit parameters of the linear regression of ERC as a function of cycling for first-reaction order (phosphate, carbonate and arsenate electrolyte) and zero-reaction order (borate electrolyte). The first reaction order is fitted with the equation: $ln\left(\frac{ERC_{end}-ERC_i}{ERC_{end}}\right)$ and the zero-reaction order with: $\frac{ERC_{end}-ERC_i}{ERC_{end}}$. The graphs are shown in Figure 8.

| ***Parameter*** | **First-reaction order** | | | **Zero-reaction order** |
|---|---|---|---|---|
| | Phosphate | Carbonate | Arsenate | Borate |
| *Slope* | -0.0046 ± 0.0002 | -0.0033 ± 0.0001 | -0.0033 ± 0.0002 | -0.0044 ± 0.0001 |
| *Intercept* | -0.11 ± 0.02 | -0.053 ± 0.007 | -0.17 ± 0.03 | 0.996 ± 0.007 |
| $R^2$ | 0.994 | 0.994 | 0.985 | 0.998 |

**Table S3.** Co nominal oxidation state of Ery in phosphate and borate electrolyte in selected cycles. The fit equation and graph are shown in Fig. S7.

| **Electrolyte** | **Co nominal oxidation state in a selected cycle** | | | |
|---|---|---|---|---|
| | 25 | 100 | 400 | 800 |
| Borate | 2.3 | 2.6 | 2.8 | 2.8 |
| Phosphate | 2.3 | 2.4 | 2.4 | 2.4 |
| Carbonate | - | 2.4 | - | 2.8 |

**Table S4.** EXAFS absorber–backscatter distance (R) and coordination numbers (N) as determined by simulation of the $k^3$-weighted EXAFS spectra at the cobalt K-edge for samples Ery-BO3, Ery-PO4 and Ery-CO3, after a selected number of cycles. The Debye-Waller factor (σ) was fixed at a value of $2\sigma^2$=0.01, as used previously for similar materials.[2]

| *Sample* | *Co-O* | *Co-O* | *Co-O* | *Co-Co* | *Co-Co* | *R-factor* |
|---|---|---|---|---|---|---|
| *Ery-BO3, after 25 cycles* | | | | | | |
| *R* | 1.836 | 2.0(1) | 2.11(6) | 2.81(1) | 3.04(1) | 10.5 |
| *N* | 1.4(6) | 3(1) | 3(2) | 2.6(3) | 2.3(3) | |
| *Ery-BO3, after 100 cycles* | | | | | | |
| *R* | 1.92(2) | 2.25(2) | 2.06(5) | 2.85(1) | 3.09(1) | 11.3 |
| *N* | 3.3(7) | 2.7(5) | 2.0(6) | 4.7(3) | 3.3(3) | |
| *Ery-BO3, after 400 cycles* | | | | | | |
| *R* | 1.91(1) | - | - | 2.82(1) | - | 11.0 |
| *N* | 4.4(2) | - | - | 4.0(2) | - | |
| *Ery-BO3, after 800 cycles* | | | | | | |
| *R* | 1.90(1) | - | - | 2.83(1) | - | 9.1 |
| *N* | 4.4(2) | - | - | 4.0(2) | - | |
| *Ery-PO4, after 25 cycles* | | | | | | |
| *R* | 1.92(7) | 2.10(2) | - | 2.85(4) | 3.03(1) | 5.2 |
| *N* | 2(2) | 2.11(1) | - | 0.6(4) | 1.9(4) | |
| *Ery-PO4, after 100 cycles* | | | | | | |
| *R* | 1.90(2) | 2.07(2) | 2.25(4) | 2.84(1) | 3.07(1) | 15.8 |
| *N* | 2.0(4) | 3.8(5) | 1.6(6) | 2.8(3) | 2.4(3) | |
| *Ery-PO4, after 400 cycles* | | | | | | |
| *R* | 1.90(3) | 2.06(2) | 2.25(4) | 2.83(1) | 3.05(1) | 10.6 |
| *N* | 1.7(4) | 3.2(4) | 1.3(5) | 3.2(3) | 2.5(3) | |
| *Ery-PO4, after 800 cycles* | | | | | | |
| *R* | 1.90(4) | 2.04(6) | 2.19(9) | 2.83(1) | 3.07(1) | 6.5 |
| *N* | 1.9(8) | 2.7(6) | 1.2(8) | 2.2(3) | 1.7(3) | |
| *Ery-CO3, after 100 cycles* | | | | | | |
| *R* | 1.97(6) | 2.12(2) | 2.34(4) | 2.87(3) | 3.05(2) | 5.9 |
| *N* | 1.5(6) | 5.4(6) | 1.2(8) | 0.6(3) | 1.3(4) | |
| *Ery-CO3, after 800 cycles* | | | | | | |
| *R* | 1.89(1) | - | - | 2.81(1) | - | 7.0 |
| *N* | 4.0(2) | - | - | 3.3(2) | - | |

The R-factor used Fourier filtered data between 1 and 3 Å using the formula $R_f = 100\,\frac{\Sigma\left(m_i^{ff}-e_i^{ff}\right)^2}{\Sigma\left(e_i^{ff}\right)^2}$, where $m^{ff}$ represents the Fourier-filtered model and $e^{ff}$ represents the experimental k-weighted EXAFS curve.

**Table S6.** Catalytic properties of common Co-based OER catalysts and the $Mn_4Ca$ oxygen evolution cluster in photosystem II. The turnover frequency (TOF) values are reported at specific overpotentials and conditions, references (Ref.) are shown.

| *Active material* | *Conditions* | *pH* | *Overpotential (mV)* | *TOF ($O_2$ $Co^{-1}$ $s^{-1}$)* | *Method* | *Ref.* |
|---|---|---|---|---|---|---|
| *Ery-CO3, after 800 cycles* | Carbonate electrolyte 0.1 M | 7 | 300 | $1.3x10^{-2}$ | ERC[a] | This work |
| | | | 310 | $1.3x10^{-2}$ | ERC[a] | |
| | | | 320 | $1.4x10^{-2}$ | ERC[a] | |
| | | | 350 | $1.6x10^{-2}$ | ERC[a] | |
| | | | 370 | $1.6x10^{-2}$ | ERC[a] | |
| | | | 400 | $1.7x10^{-2}$ | ERC[a] | |
| | | | 410 | $1.7x10^{-2}$ | ERC[a] | |
| | | | 440 | $2.0x10^{-2}$ | ERC[a] | |
| | | | 450 | $2.2\ x10^{-2}$ | ERC[a] | |
| | | | 470 | $2.3\ x10^{-2}$ | ERC[a] | |
| | | | 620 | $3.7\ x10^{-2}$ | ERC[a] | |
| | | | 870 | $1.1\ x10^{-1}$ | ERC[a] | |
| *Reconstructed Pakhomovskyite* | KPi 0.1 M | 7 | 620 | ≈0.1 | ERC[a] | [3] |
| *Co-Pi* | KPi 0.1 M | 7 | 410 | $2.6x10^{-3}$ | TC[c] | [4] |
| *$Co(PO_3)_2$ nanoparticles* | KPi 0.1 M | 6.4 | 440 | 0.10-0.21 | CS[b] | [5] |
| *$Co_3O_4$* | KOH 0.1 M | 13 | 320 | $\approx 2x10^{-3}$ | CS[b] | [6] |
| *$LaCoO_3$* | KOH 0.1 M | 13 | 370 | $2.5x10^{-2}$ | CS[b] | [6] |
| *Co-Pi* | KOH 0.1 M | 13 | 370 | $1x10^{-3}$-0.3 | CS[b] | [7] |
| *$Ba_{0.5}Sr_{0.5}Co_{0.8}Fe_{0.2}O_{3-\delta}$* | KOH 0.1 M | 13 | 470 | $4x10^{-3}$-0.3 | CS[b] | [8] |
| *$PrBaCo_2O_{5+\delta}$* | KOH 0.1 M | 13 | 370 | ≈2 | CS[b] | [6] |
| *$CoO_xH_y$* | KOH 1 M | 14 | 350 | $6x10^{-3}$ | CS[b] | [9] |
| *$LiMn_2O_4$* | NaOH 1 M | 14 | 450 | 0.07 | CS[b] | [10] |
| *$NiFeO_x$* | Borate electrolyte 0.1 M | 9.2 | 400 | 0.32 | ERC[a] | [11] |
| *$NiFeO_x$* | KPi 0.1 M | 7 | 400 | 0.01 | ERC[a] | [11] |
| *$Mn_4Ca$ oxygen evolution cluster in photosystem II* | Membranes particles in lumen | ≈5.5 | 300 | ≈100 | R[d] | [12,13] |

[a] TOF = i/ERC/4; [b] TOF = $i/\rho_{surface}/4$ (surface atoms from crystal structure); [c] TOF = $i/q_T/4$ (total electrodeposited charge); [d] TOF = reciprocal of total duration for one catalytic turnover.